\documentclass[prd, 11pt, superscriptaddress, nofootinbib, floatfix,preprintnumbers
]{revtex4-2} 

\usepackage{amsmath,amssymb}

\usepackage{graphicx}
\usepackage[caption=false]{subfig}
\usepackage{dcolumn}
\usepackage{bm}
\usepackage{natbib}
\usepackage{url}
\usepackage{tikz}
\usetikzlibrary{shapes}
\usetikzlibrary{positioning}
\usepackage{qcircuit}
\usepackage{braket}
\usepackage{bbm}
\usepackage{esvect}
\usepackage{ulem}
\usepackage[colorlinks=true,backref=false, linktocpage=true,
citecolor=blue,urlcolor=blue,linkcolor=blue,pdfpagemode=UseOutlines]{hyperref}
\usepackage{todonotes}
\usepackage{multirow}
\usepackage[section]{placeins}
\usepackage{booktabs}
\usepackage{appendix}

\def\simgt{\stackrel{>}{{}_\sim}}

\begin{document}


\title{Enhancing quantum utility: simulating large-scale quantum spin chains on superconducting quantum computers}

\newcommand*{\KU}{Department of Physics and Astronomy, University of Kansas, Lawrence, Kansas 66045, USA.}\affiliation{\KU} 
\newcommand*{\DU}{Department of Physics, University of Dhaka, P.O. Box 1000, Dhaka, Bangladesh.}\affiliation{\DU} 
\newcommand*{\CSIBNL}{Computational Science Initiative, Brookhaven National Laboratory, Upton, New York 11973, USA.}\affiliation{\CSIBNL} 
\newcommand*{\UAL}{Department of Physics and Astronomy, University of Alabama, Tuscaloosa, 35487, Alabama, USA.}\affiliation{\UAL} 
\newcommand*{\CBNL}{Collider-Accelerator Department, Brookhaven National Laboratory, Upton, New York 11973, USA.}\affiliation{\CBNL} 
\newcommand*{\RBRC}{RIKEN-BNL Research Center, Brookhaven National Laboratory, Upton, New York 11973, USA.}\affiliation{\RBRC} 
\newcommand*{\POBNL}{Physics Department, Brookhaven National Laboratory, Upton, New York 11973, USA.}\affiliation{\POBNL}

\author{Talal Ahmed Chowdhury}\email{talal@ku.edu}\affiliation{\KU}\affiliation{\DU}
\author{Kwangmin Yu}\email{kyu@bnl.gov}\affiliation{\CSIBNL}
\author{Mahmud Ashraf Shamim}\affiliation{\UAL}
\author{M.L. Kabir}\affiliation{\CBNL}
\author{Raza Sabbir Sufian}\affiliation{\RBRC}\affiliation{\POBNL}

\begin{abstract}
We present the quantum simulation of the frustrated quantum spin-$\frac{1}{2}$ antiferromagnetic Heisenberg spin chain with competing nearest-neighbor $(J_1)$ and next-nearest-neighbor $(J_2)$ exchange interactions in the real superconducting quantum computer with qubits ranging up to 100. In particular, we implement, for the first time, the Hamiltonian with the next-nearest neighbor exchange interaction in conjunction with the nearest neighbor interaction on IBM's superconducting quantum computer and carry out the time evolution of the spin chain by employing the first-order Trotterization. Furthermore, our novel implementation of the second-order Trotterization for the isotropic Heisenberg spin chain, involving only nearest-neighbor exchange interaction, enables precise measurement of the expectation values of staggered magnetization observable across a range of up to 100 qubits. Notably, in both cases, our approach results in a constant circuit depth in each Trotter step, independent of the number of qubits. Our demonstration of the accurate measurement of expectation values for the large-scale quantum system using superconducting quantum computers designates the quantum utility of these devices for investigating various properties of many-body quantum systems. This will be a stepping stone to achieving the quantum advantage over classical ones in simulating quantum systems before the fault tolerance quantum era.  
 \end{abstract}

\maketitle

\newpage
\section{Introduction}\label{sec:intro}

The landscape of quantum computing has experienced significant evolution, especially with the emergence of noisy intermediate-scale quantum (NISQ) computers~\cite{preskill_NISQ, Bharti:2021zez} and beyond at scale such as IBM Quantum processors. Despite their inherent noise and limitations, these platforms have opened up new avenues for delving into fundamental physics. Quantum simulation~\cite{cirac-zollar, Georgescu:2013oza, Daley:2022eja} of seemingly complex many-body quantum systems using near-term, noisy quantum computers presents an intriguing possibility. While the algorithm for quantum simulation using quantum computers was initially outlined for many-body Hamiltonians in~\cite{Lloyd} and subsequently refined in works such as~\cite{Lloyd-Abrams, somaroo, zalka, farhi, ortiz, somma-ortiz, berry, childs}, its actual implementation on a quantum computer necessitates comprehensive quantum error correction.


Using near-term, noisy quantum computers to simulate fundamental physics presents significant challenges including error rates that affect computation accuracy, constraints on qubit numbers limiting the complexity of simulated systems, and difficulty in maintaining qubit stability over extended periods. Nevertheless, ongoing advancements in error-mitigation techniques and algorithms~\cite{endo-error-mitigation, Temme-error-mitigation, Li-error-mitigation, Kandala-error-mitigation, Berg-error-mitigation, yu2023simulating, Kim-error-mitigation, kim2023evidence} for noisy quantum devices are enhancing their capability to perform detailed and accurate simulations of fundamental physics. These successes showed the utility of noisy quantum computers before the advent of fault-tolerance~\cite{kim2023evidence}. Despite these advancements, an important question remains: are currently available quantum computers capable of simulating large quantum systems and extracting precise values for observables on more realistic problems? This question warrants further investigation to assess the practical limitations and potential of current quantum technology in the realm of large-scale quantum simulations. Recently,  Kim $\textit{et al}$. \cite{kim2023evidence} successfully performed time evolution simulation of the Ising model on IBM quantum computers at a scale beyond exact classical methods with accuracy competitive with tensor network methods. However, it remains an open problem to achieve such quantum utility for a broader range of practical problems. 

In this study, we expand the utility of noisy quantum computers to more general and complicated cases of time evolution driven by Hamiltonians at large-scale noisy superconducting quantum computers. We focus on the simulation for the time evolution of quantum spin-$\frac{1}{2}$ antiferromagnetic Heisenberg model with frustration and assess their ability to accurately capture the intricate spin dynamics of the model. The frustrated spin-$\frac{1}{2}$ antiferromagnetic model serves as a paradigmatic representation of a quantum many-body system characterized by competing interactions among its constituents. In a magnetically frustrated system, the ground state becomes degenerate due to the inherent ambiguity of the spin configurations not being able to satisfy all of the antiferromagnetic interactions simultaneously. Consequently, the ground state of the frustrated systems becomes highly entangled, leading to exotic phases of quantum matters such as Quantum Spin Liquids (QSL)~\cite{Fazekas1974, Shastry-Sutherland, Balents2010SpinLI}.

In particular, we consider the spin-$\frac{1}{2}$ antiferromagnetic spin chain with competing nearest-neighbor $(J_1)$ and next-nearest-neighbor $(J_2)$ exchange interactions~\cite{Haldane-1, Haldane-2} in the real superconducting quantum computer with qubits ranging up to 100. 
The antiferromagnetic quantum spin chain has a rich ground state quantum phase diagram~\cite{Okamoto, Nomura_1994, white-affleck, eggert, murdy}. Apart from its rich quantum phase structure, interestingly, the antiferromagnetic spin chain model can be related to the Schwinger model~\cite{Wiegmann:1991zy, Diamantini:1992ei, Hosotani:1996sn, Hosotani:1997kv}, a toy model in 1+1 D that captures the features of a strongly coupled sector of QCD.
In order to simulate the time evolution of the quantum spin-$\frac{1}{2}$ antiferromagnetic Heisenberg spin chain with competing nearest-neighbor and next-nearest-neighbor exchange interactions on noisy superconducting quantum computers, we developed a new Trotterization \cite{trotter, suzuki1, suzuki2} circuit design. The main challenge of the circuit design is originated by the limited connectivity of superconducting quantum computers. Since the model has the interaction between the next-nearest-neighbor in addition to the nearest neighbor, the limited connectivity of the system is a huge barrier for efficient Trotterization while the nearest-neighbor interaction can be efficiently implemented on linear qubit connectivity. 
Our new circuit design for the model is suitable for linear qubit connectivity (circular connectivity for a periodic boundary condition).
Also, the design has a constant circuit depth with respect to the system size (the number of qubits) so that this implementation is scalable.
This circuit design is described in detail in Sec. \ref{sec:time-evoliton_XXX_XXX2}.

Moreover, a special case ($J_2$ = 0) of the model is the Heisenberg isotropic spin chain model. In this case, we propose a new second-order Trotterization implementation.
In general, a second-order Trotterization has twice the longer circuit depth than the corresponding first-order Trotterization.
However, we achieve the second-order Trotterization by only an additional constant circuit depth than the circuit depth of the first-order Trottorization.
Since we have a trade-off between numerical noise and quantum device noise when we increase the order of Trotterization, implementing the second-order Trotterization with only constant circuit depth increase from the first-order Trotterization is a great benefit. 
The implementation detail is described in Sec. \ref{sec:time-evoliton_XXX}.

Subsequently, we validate our new circuit designs with 20, 96, and 100 qubit systems on the IBM quantum processors of 127 qubits and 133 qubits.
To cope with the quantum errors and noises, we apply several quantum error mitigation methods to our new circuit designs (Refer to Sec. \ref{sec:error_mitigations}).
We successfully simulate the time evolution with 3888 and 3978 $\texttt{CX}$ gates using open and periodic boundary conditions, respectively, that are presented in Sec.~\ref{sec:results}. Finally, we conclude in Sec.~\ref{sec:conclusion}.

\section{Frustrated spin-$\frac{1}{2}$ antiferromagnetic spin chain model}\label{sec:model}
The frustrated spin-$\frac{1}{2}$ antiferromagnetic Heisenberg spin chain is described by the following Hamiltonian,
\begin{align}
    H &= J_{1}\sum_{i=1}^{N}\left(S^{x}_{i}S^{x}_{i+1}+S^{y}_{i}S^{y}_{i+1}+\Delta S^{z}_{i}S^{z}_{i+1}\right)+ J_{2}\sum_{i=1}^{N}\left(S^{x}_{i}S^{x}_{i+2}+S^{y}_{i}S^{y}_{i+2}+ S^{z}_{i}S^{z}_{i+2}\right),\label{eq:hamiltonian}
\end{align}
where the antiferromagnetic nearest-neighbor (NN) coupling $J_{1}>0$, next-nearest-neighbor (NNN) coupling $J_{2}\geq 0$ and the exchange-anisotropy parameter $\Delta\geq 0$ control the parameter space of the Hamiltonian. Besides, the spin operators, $S^{i}=\frac{1}{2}\sigma^{i}$ obey the $SU(2)$ algebra,
\begin{equation}
[S^{\alpha}_{i},S^{\beta}_{j}]=i\delta_{ij}\epsilon^{\alpha\beta\gamma}S^{\gamma}_{i},
\end{equation}
where $\alpha,\,\beta,\,\gamma=x,\,y,\,z$ and $i,\,j=1,...,N$. Our analysis considers open boundary conditions (OBC) and periodic boundary conditions (PBC). The PBC is imposed by setting $S^{\alpha}_{i+N}=S^{\alpha}_{i}$. Besides, we take the total number of spin sites on the chain as even $N=4n$, where $n$ takes on positive integers. Additionally, this Hamiltonian is referred to spin-$\frac{1}{2}$ $J_{1}-J_{2}$ XXZ Hamiltonian in many instances.
\begin{figure}[h!]
\centering
\includegraphics[width=0.55\textwidth]{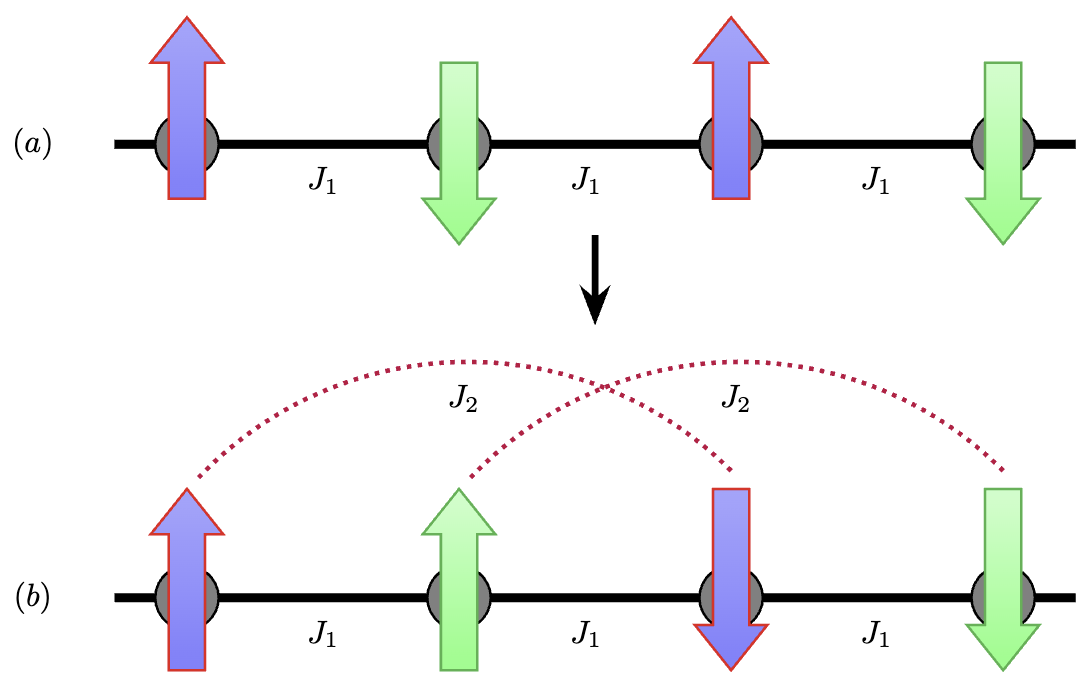}
\caption{(a) In the absence of frustration, the nearest-neighbor interaction prefers the antiferromagnetic or the N\'eel ordering. (b) In contrast, the onset of next-nearest-neighbor interaction $J_{2}$ makes the system frustrated as it favors the anti-parallel alignment of the next-nearest-neighboring spins, leading to a parallel combination between neighboring spin pairs.}
\label{fig: Frustrated spin chain}
\end{figure}
From Fig. \ref{fig: Frustrated spin chain}, we can see that when the next-nearest-neighbor interaction $J_{2}$ is set to zero, the spin alignment, for example along the $z$-axis, follows the antiferromagnetic ordering but the onset of $J_{2}$ introduces a competing interaction which would disrupt the initial antiferromagnetic ordering for large enough $J_{2}$ value. Hence, the spin chain becomes frustrated.

The couplings $(J_{1},\,J_{2},\,\Delta)$ of the Hamiltonian would result in a rich ground state phase diagram of the frustrated quantum spin-$\frac{1}{2}$ antiferromagnetic spin chain. In the subsequent analysis, we focus on two important Hamiltonians for particular parameter values, as detailed below.

\noindent\textbf{Isotropic Heisenberg Hamiltonian:} The isotropic Heisenberg Hamiltonian, also known $H_{\mathrm{iso}}$, is characterized by the parameters, $J_{1}>0$, $J_{2}=0$ and $\Delta=1$. Unlike the general case of $\Delta\neq 1$, it has a full global $SU(2)$ symmetry.

\noindent\textbf{Dimer Hamiltonian:} The Dimer Hamiltonian corresponding to the Majumder-Ghosh (MG) point, denoted here as $H_{\mathrm{Dimer}}$, is characterized by $J_{1}>0,\,J_{2}=\frac{J_{1}}{2},\Delta=1$~\cite{majumdar-1, majumdar-2}. It also enjoys the full $SU(2)$ symmetry. The crucial feature of this Hamiltonian is that its ground state manifests as a doubly degenerate valence bond solid (VBS) phase where the pairs of neighboring spins on the chain form spin-singlets, referred to as Dimer states. 

\subsection{Time evolution of the quantum system}
In this work, we focus on the time evolution of the spin chain under the Hamiltonians, $H_{\mathrm{iso}}$ and $H_{\mathrm{Dimer}}$. As our focus is to study the accuracy of the measurement of observables associated with the spin chain at the superconducting quantum computers, we focus on the temporal variation of the expectation value of the staggered magnetization that characterizes the antiferromagnetic ordering in the spin chain. The staggered magnetization observable $\hat{O}_{M_{st}}$ is defined as follows:
\begin{equation}
    \hat{O}_{M_{st}}=\frac{1}{N}\sum_{i=1}^{N}(-1)^{i}S^{z}_{i} .\label{staggered-mag}
\end{equation}
One can choose a specific spin configuration of the quantum spin chain and calculate the expectation value of the staggered magnetization observable to characterize the spin states' antiferromagnetic ordering. There exist myriad options for selecting such spin states. However, for simplicity and clarity, we opt for the N\'eel state that encapsulates some of the fundamental features of the antiferromagnetic spin chain. It is defined as,
\begin{equation}
    |\psi_{\mathrm{Neel}}\rangle = |\uparrow\downarrow\uparrow\downarrow \cdots \cdots \uparrow\downarrow\uparrow\downarrow\rangle\,,
    \label{eq: neel-state}
\end{equation}
where each $|\uparrow\rangle$ or $|\downarrow\rangle$ represent the spin projection of spin-$1/2$ particle at `i'-th site along the $z$-axis in spin space.

Consequently, we determine the time evolution of the expectation value of staggered magnetization observable for the N\'eel state under the Hamiltonian $H_{\mathrm{iso}}$ and $H_{\mathrm{Dimer}}$ in IBM's superconducting quantum computers and corroborated the results with state-of-the-art classical numerical tools.

\section{Implementation of time evolution under the spin chain Hamiltonian}\label{sec:time-evoliton}

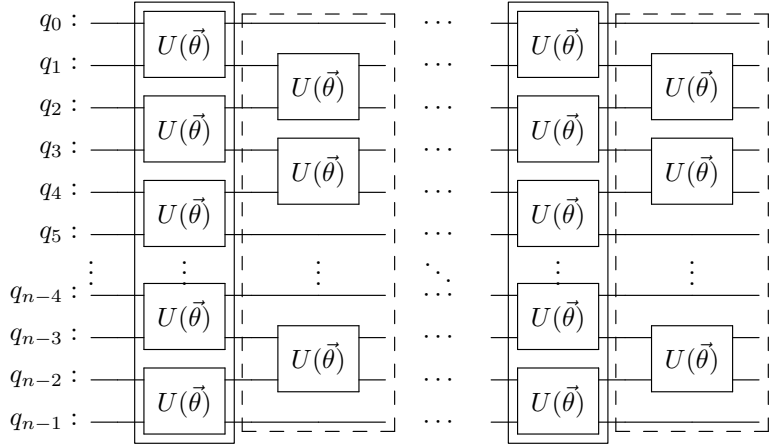
\begin{figure}[t!]
\[
\Qcircuit @C=1em @R=0.7em {
\lstick{ {q}_{0} :  } & \qw & \multigate{1}{U (\vec{\theta})} & \qw & \qw & \qw & & \cdots & & & \multigate{1}{U (\vec{\theta})} & \qw & \qw & \qw  \\
\lstick{ {q}_{1} :  } & \qw & \ghost{U (\vec{\theta})} & \qw  & \multigate{1}{U (\vec{\theta})} & \qw & & \cdots & & & \ghost{U (\vec{\theta})} & \qw  & \multigate{1}{U (\vec{\theta})} & \qw \\
\lstick{ {q}_{2} :  } & \qw & \multigate{1}{U (\vec{\theta})} & \qw  & \ghost{U (\vec{\theta})} & \qw & & \cdots & & & \multigate{1}{U (\vec{\theta})} & \qw  & \ghost{U (\vec{\theta})} & \qw \\
\lstick{ {q}_{3} :  } & \qw & \ghost{U (\vec{\theta})} & \qw & \multigate{1}{U (\vec{\theta})} & \qw & & \cdots&  & & \ghost{U (\vec{\theta})} & \qw & \multigate{1}{U (\vec{\theta})} & \qw \\
\lstick{ {q}_{4} :  } & \qw & \multigate{1}{U (\vec{\theta})} & \qw & \ghost{U (\vec{\theta})} & \qw & & \cdots & & & \multigate{1}{U (\vec{\theta})} & \qw & \ghost{U (\vec{\theta})} & \qw \\
\lstick{ {q}_{5} :  } & \qw & \ghost{U (\vec{\theta})}& \qw & \qw & \qw & & \cdots & & & \ghost{U (\vec{\theta})}& \qw & \qw & \qw \\
\vdots  &  & \vdots & & \vdots & & &\ddots & & & \vdots & & \vdots \\
\lstick{ {q}_{n-4} :  } & \qw & \multigate{1}{U (\vec{\theta})} & \qw & \qw & \qw & & \cdots & & & \multigate{1}{U (\vec{\theta})} & \qw & \qw & \qw \\
\lstick{ {q}_{n-3} :  } & \qw & \ghost{U (\vec{\theta})}& \qw & \multigate{1}{U (\vec{\theta})} & \qw & & \cdots & & & \ghost{U (\vec{\theta})}& \qw & \multigate{1}{U (\vec{\theta})} & \qw \\
\lstick{ {q}_{n-2} :  } & \qw & \multigate{1}{U (\vec{\theta})} & \qw & \ghost{U (\vec{\theta})} & \qw & & \cdots & & & \multigate{1}{U (\vec{\theta})} & \qw & \ghost{U (\vec{\theta})} & \qw \\
\lstick{ {q}_{n-1} :  } & \qw & \ghost{U (\vec{\theta})}& \qw & \qw & \qw & & \cdots & & & \ghost{U (\vec{\theta})}& \qw & \qw & \qw
\gategroup{1}{3}{11}{3}{.7em}{-}
\gategroup{1}{4}{11}{6}{.7em}{--}
\gategroup{1}{11}{11}{11}{.7em}{-}
\gategroup{1}{12}{11}{14}{.7em}{--}
}
\]
\vspace*{1mm}
\caption{The first-order Trotterization of the Hamiltonian $H_{\mathrm{iso}}$ with open boundary condition. The layers surrounded by the straight lines are the even layers and the layers surrounded by the dotted lines are the odd layers. For a periodic boundary condition, the odd layers have the two-qubit gates, $U(\vec{\theta})$, between $q_{n-1}$ and $q_0$. One Trotter step is composed of the even layer and the odd layer.}
\label{fig:circuit_XXX_1st}
\end{figure}

Starting from this section, we use $0$ as the first index instead of $1$ to keep consistency with IBM Qiskit's qubit index convention.
Hence, the index varies from $0$ to $N-1$ instead of from $1$ to $N$.
Also, we assume $N$ is even.
Eq.~\eqref{eq:hamiltonian} is reformulated by the Pauli operators, $\sigma^x_j,\sigma^y_j,\sigma^z_j$ as follows:
\begin{equation}\label{eq:hamiltonian_Pauli}
H = \sum_{j=0}^{N-1} \Bigl(J_x\sigma_j^x\sigma_{j+1}^x+J_y\sigma_j^y\sigma_{j+1}^y+J_z\sigma_j^z\sigma_{j+1}^z \Bigr) + 
\frac{J_2}{4} \sum_{j=0}^{N-1} \Bigl( \sigma_j^x\sigma_{j+2}^x + \sigma_j^y\sigma_{j+2}^y + \sigma_j^z\sigma_{j+2}^z \Bigr) ,
\end{equation}
where $J_z = \Delta J_1 / 4$ and $J_x = J_y = J_1 / 4$.

\subsection{The second-order Trotterization for isotropic Heisenberg Hamiltonian}
\label{sec:time-evoliton_XXX}

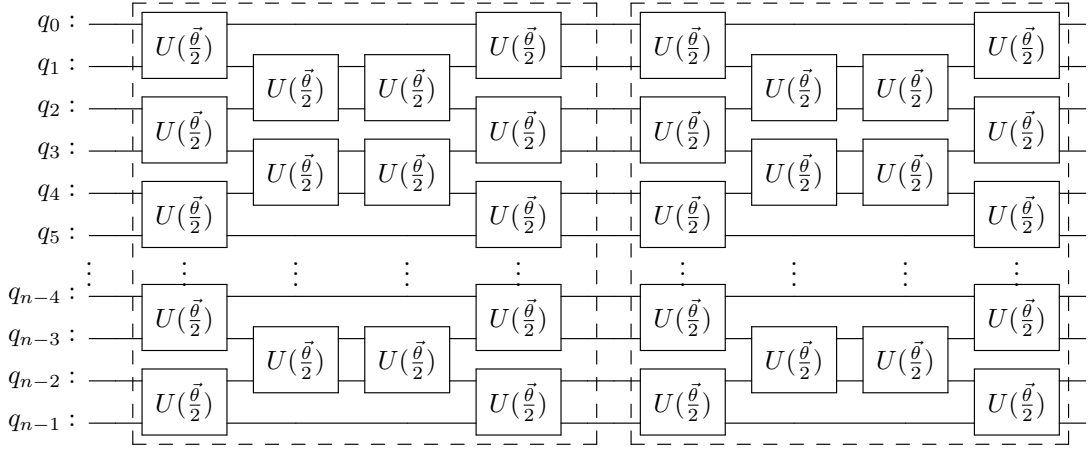
\begin{figure}[t!]
\[
\Qcircuit @C=1em @R=0.7em {
\lstick{ {q}_{0} :  }   & \qw & \multigate{1}{U (\frac{\vec{\theta}}{2})} & \qw & \qw & \multigate{1}{U (\frac{\vec{\theta}}{2})} & \qw   & \qw & \multigate{1}{U (\frac{\vec{\theta}}{2})} & \qw & \qw & \multigate{1}{U (\frac{\vec{\theta}}{2})} & \qw \\
\lstick{ {q}_{1} :  }   & \qw & \ghost{U (\frac{\vec{\theta}}{2})} & \multigate{1}{U (\frac{\vec{\theta}}{2})} & \multigate{1}{U (\frac{\vec{\theta}}{2})} & \ghost{U (\frac{\vec{\theta}}{2})} & \qw    & \qw & \ghost{U (\frac{\vec{\theta}}{2})} &  \multigate{1}{U (\frac{\vec{\theta}}{2})} & \multigate{1}{U (\frac{\vec{\theta}}{2})} & \ghost{U (\frac{\vec{\theta}}{2})} & \qw   \\
\lstick{ {q}_{2} :  }   & \qw & \multigate{1}{U (\frac{\vec{\theta}}{2})} & \ghost{U (\frac{\vec{\theta}}{2})}  & \ghost{U (\frac{\vec{\theta}}{2})} &  \multigate{1}{U (\frac{\vec{\theta}}{2})} & \qw  & \qw & \multigate{1}{U (\frac{\vec{\theta}}{2})}   & \ghost{U (\frac{\vec{\theta}}{2})}  & \ghost{U (\frac{\vec{\theta}}{2})} & \multigate{1}{U (\frac{\vec{\theta}}{2})} & \qw \\
\lstick{ {q}_{3} :  }   & \qw & \ghost{U (\frac{\vec{\theta}}{2})} & \multigate{1}{U (\frac{\vec{\theta}}{2})} & \multigate{1}{U (\frac{\vec{\theta}}{2})} &  \ghost{U (\frac{\vec{\theta}}{2})} & \qw    & \qw & \ghost{U (\frac{\vec{\theta}}{2})} & \multigate{1}{U (\frac{\vec{\theta}}{2})} & \multigate{1}{U (\frac{\vec{\theta}}{2})} & \ghost{U (\frac{\vec{\theta}}{2})} & \qw   \\
\lstick{ {q}_{4} :  }   & \qw & \multigate{1}{U (\frac{\vec{\theta}}{2})} & \ghost{U (\frac{\vec{\theta}}{2})} & \ghost{U (\frac{\vec{\theta}}{2})} &  \multigate{1}{U (\frac{\vec{\theta}}{2})} & \qw    & \qw & \multigate{1}{U (\frac{\vec{\theta}}{2})} & \ghost{U (\frac{\vec{\theta}}{2})} & \ghost{U (\frac{\vec{\theta}}{2})} & \multigate{1}{U (\frac{\vec{\theta}}{2})} & \qw   \\
\lstick{ {q}_{5} :  }   & \qw & \ghost{U (\frac{\vec{\theta}}{2})} & \qw & \qw & \ghost{U (\frac{\vec{\theta}}{2})}& \qw & \qw & \ghost{U (\frac{\vec{\theta}}{2})} & \qw & \qw & \ghost{U (\frac{\vec{\theta}}{2})}& \qw \\
\vdots  &  & \vdots & \vdots & \vdots & \vdots & & & \vdots & \vdots & \vdots & \vdots \\
\lstick{ {q}_{n-4} :  } & \qw & \multigate{1}{U (\frac{\vec{\theta}}{2})} & \qw & \qw & \multigate{1}{U (\frac{\vec{\theta}}{2})} & \qw                                                    & \qw & \multigate{1}{U (\frac{\vec{\theta}}{2})} & \qw & \qw & \multigate{1}{U (\frac{\vec{\theta}}{2})} & \qw &                                                   \\
\lstick{ {q}_{n-3} :  } & \qw & \ghost{U (\frac{\vec{\theta}}{2})}& \multigate{1}{U (\frac{\vec{\theta}}{2})} & \multigate{1}{U (\frac{\vec{\theta}}{2})} & \ghost{U (\frac{\vec{\theta}}{2})}& \qw      & \qw & \ghost{U (\frac{\vec{\theta}}{2})}& \multigate{1}{U (\frac{\vec{\theta}}{2})} & \multigate{1}{U (\frac{\vec{\theta}}{2})} & \ghost{U (\frac{\vec{\theta}}{2})}& \qw     \\
\lstick{ {q}_{n-2} :  } & \qw & \multigate{1}{U (\frac{\vec{\theta}}{2})} & \ghost{U (\frac{\vec{\theta}}{2})} & \ghost{U (\frac{\vec{\theta}}{2})} & \multigate{1}{U (\frac{\vec{\theta}}{2})} & \qw    & \qw & \multigate{1}{U (\frac{\vec{\theta}}{2})} & \ghost{U (\frac{\vec{\theta}}{2})} & \ghost{U (\frac{\vec{\theta}}{2})} & \multigate{1}{U (\frac{\vec{\theta}}{2})} & \qw   \\
\lstick{ {q}_{n-1} :  } & \qw & \ghost{U (\frac{\vec{\theta}}{2})} & \qw & \qw & \ghost{U (\frac{\vec{\theta}}{2})}& \qw                                                                      & \qw & \ghost{U (\frac{\vec{\theta}}{2})} & \qw & \qw &  \ghost{U (\frac{\vec{\theta}}{2})}& \qw 
\gategroup{1}{3}{11}{7}{.7em}{--}
\gategroup{1}{9}{11}{12}{.7em}{--}
}
\]
\vspace*{1mm}
\caption{The second-order Trotterization of the Hamiltonian $H_{\mathrm{iso}}$ with open boundary condition with two Trotter steps. The dotted parts are one Trotter step of the second-order. The one step of the second-order Trotterizatoin is composed of an even layer, an odd layer, an odd layer, and an even layer in order. For a periodic boundary condition, the odd layers have the two-qubit gates, $U(\vec{\frac{\theta}{2}})$, between $q_{n-1}$ and $q_0$.}
\label{fig:circuit_XXX_2nd}
\end{figure}

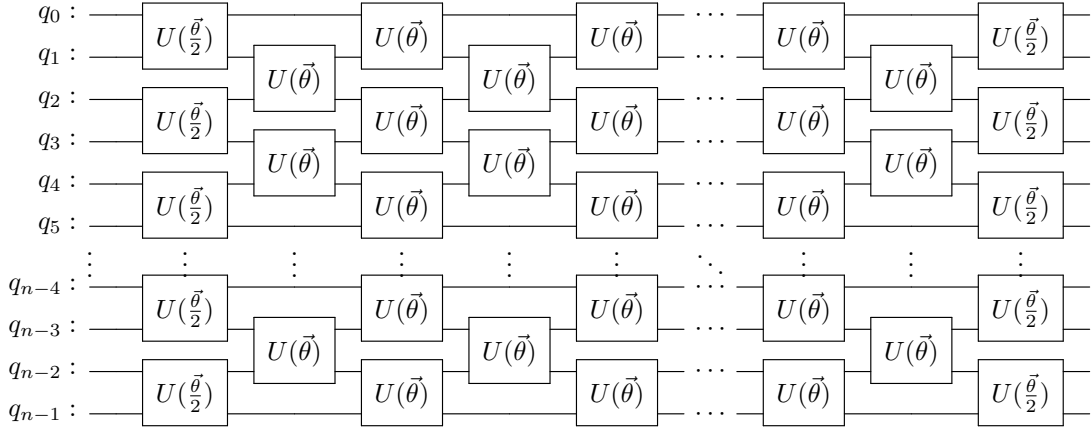
\begin{figure}[t!]
\[
\Qcircuit @C=1em @R=0.7em {
\lstick{ {q}_{0} :  }   & \qw & \multigate{1}{U (\frac{\vec{\theta}}{2})} & \qw & \multigate{1}{U (\vec{\theta})} & \qw & \multigate{1}{U (\vec{\theta})} & \qw & \cdots & & \multigate{1}{U (\vec{\theta})} & \qw & \multigate{1}{U (\frac{\vec{\theta}}{2})} & \qw \\
\lstick{ {q}_{1} :  }   & \qw & \ghost{U (\frac{\vec{\theta}}{2})} & \multigate{1}{U (\vec{\theta})} & \ghost{U (\vec{\theta})} & \multigate{1}{U (\vec{\theta})} & \ghost{U (\vec{\theta})} & \qw & \cdots & & \ghost{U (\vec{\theta})} & \multigate{1}{U (\vec{\theta})} & \ghost{U (\frac{\vec{\theta}}{2})} & \qw \\
\lstick{ {q}_{2} :  }   & \qw & \multigate{1}{U (\frac{\vec{\theta}}{2})} & \ghost{U (\vec{\theta})} &  \multigate{1}{U (\vec{\theta})} & \ghost{U (\vec{\theta})}  & \multigate{1}{U (\vec{\theta})} & \qw & \cdots & &  \multigate{1}{U (\vec{\theta})} & \ghost{U (\vec{\theta})}  & \multigate{1}{U (\frac{\vec{\theta}}{2})} & \qw \\
\lstick{ {q}_{3} :  }   & \qw & \ghost{U (\frac{\vec{\theta}}{2})} & \multigate{1}{U (\vec{\theta})} &  \ghost{U (\vec{\theta})} & \multigate{1}{U (\vec{\theta})} & \ghost{U (\vec{\theta})} & \qw & \cdots & &  \ghost{U (\vec{\theta})} & \multigate{1}{U (\vec{\theta})} & \ghost{U (\frac{\vec{\theta}}{2})} & \qw \\
\lstick{ {q}_{4} :  }   & \qw & \multigate{1}{U (\frac{\vec{\theta}}{2})} & \ghost{U (\vec{\theta})} &  \multigate{1}{U (\vec{\theta})} & \ghost{U (\vec{\theta})} & \multigate{1}{U (\vec{\theta})} & \qw & \cdots & &  \multigate{1}{U (\vec{\theta})} & \ghost{U (\vec{\theta})} & \multigate{1}{U (\frac{\vec{\theta}}{2})} & \qw  \\
\lstick{ {q}_{5} :  }   & \qw & \ghost{U (\frac{\vec{\theta}}{2})} & \qw & \ghost{U (\vec{\theta})} & \qw & \ghost{U (\vec{\theta})}& \qw & \cdots & & \ghost{U (\vec{\theta})} & \qw & \ghost{U (\frac{\vec{\theta}}{2})}& \qw \\
\vdots  &  & \vdots & \vdots & \vdots & \vdots & \vdots &  & \ddots & & \vdots & \vdots & \vdots &  \\
\lstick{ {q}_{n-4} :  } & \qw & \multigate{1}{U (\frac{\vec{\theta}}{2})} & \qw & \multigate{1}{U (\vec{\theta})} & \qw & \multigate{1}{U (\vec{\theta})} & \qw & \cdots & & \multigate{1}{U (\vec{\theta})} & \qw & \multigate{1}{U (\frac{\vec{\theta}}{2})} & \qw \\
\lstick{ {q}_{n-3} :  } & \qw & \ghost{U (\frac{\vec{\theta}}{2})}& \multigate{1}{U (\vec{\theta})} & \ghost{U (\vec{\theta})} & \multigate{1}{U (\vec{\theta})} & \ghost{U (\vec{\theta})}& \qw & \cdots & & \ghost{U (\vec{\theta})} & \multigate{1}{U (\vec{\theta})} & \ghost{U (\frac{\vec{\theta}}{2})}& \qw \\
\lstick{ {q}_{n-2} :  } & \qw & \multigate{1}{U (\frac{\vec{\theta}}{2})} & \ghost{U (\vec{\theta})} & \multigate{1}{U (\vec{\theta})} & \ghost{U (\vec{\theta})} & \multigate{1}{U (\vec{\theta})} & \qw  & \cdots & & \multigate{1}{U (\vec{\theta})} & \ghost{U (\vec{\theta})} & \multigate{1}{U (\frac{\vec{\theta}}{2})} & \qw \\
\lstick{ {q}_{n-1} :  } & \qw & \ghost{U (\frac{\vec{\theta}}{2})} & \qw & \ghost{U (\vec{\theta})} & \qw & \ghost{U (\vec{\theta})}& \qw & \cdots & & \ghost{U (\vec{\theta})} & \qw & \ghost{U (\frac{\vec{\theta}}{2})}& \qw
}
\]
\vspace*{1mm}
\caption{The optimized second-order Trotterization of the Hamiltonian $H_{\mathrm{iso}}$ with open boundary condition with two Trotter steps. For a periodic boundary condition, the odd layers have the two-qubit gates, $U(\vec{\theta})$, between $q_{n-1}$ and $q_0$. The circuit diagram in Fig. \ref{fig:circuit_XXX_2nd} is optimized by merging adjacent even layers and odd layers, respectively.}
\label{fig:circuit_XXX_2nd_opt}
\end{figure}

In this section, we address a specific case of the Hamiltonian (Eq.~\eqref{eq:hamiltonian_Pauli}), which has $J_2 = 0$.
We have a basic building block for the time evolution as follows:
\begin{align}
\label{eq:evolution_XYZ}
U_i (\vec{\theta}) = \exp \left(  -i \left( \frac{\theta_x}{2} \sigma_i^x \sigma_{i+1}^x + \frac{\theta_y}{2} \sigma_i^y \sigma_{i+1}^y + \frac{\theta_z}{2} \sigma_i^z \sigma_{i+1}^z \right) \right) ,
\end{align}
where $\vec{\theta} = (\theta_x, \theta_y, \theta_z) = (2 J_x \Delta t, ~ 2 J_y \Delta t, ~ 2 J_z \Delta t)$ with the Trotter step size $\Delta t$.
By the Trotter approximation, we can arrange the $U_i (\vec{\theta})$ operators in staggered placement \cite{vanicat2018integrable, smith2019simulating} as shown in Fig. \ref{fig:circuit_XXX_1st}.
Hence, one Trotter step is formulated as follows:
\begin{align}
\nonumber
U (\vec{\theta}) = \left( \prod_{j=0}^{N/2 - 1} U_{2i} (\vec{\theta}) \right) \left( \prod_{j=0}^{N/2 - 1} U_{2i+1} (\vec{\theta}) \right).
\end{align}
We define the even layer ($U_e (\vec{\theta})$) and the odd layer ($U_o (\vec{\theta})$) as follow:
\begin{align}
\nonumber
U_e (\vec{\theta}) = \left( \prod_{j=0}^{N/2 - 1} U_{2i} (\vec{\theta}) \right) , ~~
U_o (\vec{\theta}) = \left( \prod_{j=0}^{N/2 - 1} U_{2i+1} (\vec{\theta}) \right),
\end{align}
respectively.
The even layers and the odd layers are highlighted in straight lines and dotted lines, respectively in Fig. \ref{fig:circuit_XXX_1st}.
The first-order Trotterization (Fig. \ref{fig:circuit_XXX_1st}) needs $2 M$ layers when we have $M$ Trotter steps.

The second-order Trotterization is described in Fig. \ref{fig:circuit_XXX_2nd}.
Even though the accuracy of the second-order Trotterization increases, the circuit depth increases double in general.
However, we can compress the circuits of the second-order Trotterization for the Heisenberg XYZ spin chain Hamiltonian.
It is trivial since we have the following equality
\begin{align}
\nonumber
U_e (\vec{\theta_1}) U_e (\vec{\theta_2}) = U_e (\vec{\theta_1} + \vec{\theta_2}) ~~~\text{and}~~~
U_o (\vec{\theta_1}) U_o (\vec{\theta_2}) = U_o (\vec{\theta_1} + \vec{\theta_2}).
\end{align}
Hence, we can merge the adjacent odd layers, and the last even layer can be merged with the first even layer of the next Trotter step.
Fig.~\ref{fig:circuit_XXX_2nd_opt} shows the merged circuit diagram of the second-order Trotterization in Fig.~\ref{fig:circuit_XXX_2nd}.
The merged second-order Trotterization depicted in Figure.~\ref{fig:circuit_XXX_2nd_opt} shows that the implementation needs only $2M + 1$ layers when we have $M$ Trotter steps. Note that the first-order Trotterization (cf. Fig.~\ref{fig:circuit_XXX_1st}) has $2M$ layers with $M$ Trotter steps.
We achieve the second-order Trotterization by adding one even layer at the end of the first-order Trotterization and adjusting the angle parameters ($\vec{\theta}$).
Considering the trade-off between the Trotterization error and the quantum noise when we use a higher-order Trotterization, it is a great benefit to achieve the second-order Trotterization with only constant circuit depth.

\subsection{The first-order Trotterization for the Dimer Hamiltonian}
\label{sec:time-evoliton_XXX_XXX2}

\begin{figure}[h!]
\[
\Qcircuit @C=1em @R=1.0em {
\lstick{ {q}_{0} :  }   & \qw & \multigate{1}{U (\vec{\theta})} & \qw & \qw & \qw & \multigate{1}{U (\vec{\theta})} & \qw & \multigate{1}{U (\vec{\theta})} & \qw \\
\lstick{ {q}_{1} :  }   & \qw & \ghost{U (\vec{\theta})} &  \multigate{1}{U (\vec{\theta})} & \qw & \qswap & \ghost{U (\vec{\theta})} & \qswap & \ghost{U (\vec{\theta})} & \qw \\
\lstick{ {q}_{2} :  }   & \qw & \multigate{1}{U (\vec{\theta})} & \ghost{U (\vec{\theta})} & \qw & \qswap \qwx &  \multigate{1}{U (\vec{\theta})} & \qswap \qwx  & \multigate{1}{U (\vec{\theta})} & \qw \\
\lstick{ {q}_{3} :  }   & \qw & \ghost{U (\vec{\theta})} & \multigate{1}{U (\vec{\theta})} & \qw & \qw &  \ghost{U (\vec{\theta})} & \qswap & \ghost{U (\vec{\theta})} & \qswap \\
\lstick{ {q}_{4} :  }   & \qw & \multigate{1}{U (\vec{\theta})} & \ghost{U (\vec{\theta})} & \qw & \qw &  \multigate{1}{U (\vec{\theta})} & \qswap \qwx & \multigate{1}{U (\vec{\theta})} & \qswap \qwx \\
\lstick{ {q}_{5} :  }   & \qw & \ghost{U (\vec{\theta})} & \multigate{1}{U (\vec{\theta})} & \qw & \qswap &  \ghost{U (\vec{\theta})} & \qswap & \ghost{U (\vec{\theta})} & \qw \\
\lstick{ {q}_{6} :  }   & \qw & \multigate{1}{U (\vec{\theta})} & \ghost{U (\vec{\theta})} & \qw & \qswap \qwx & \multigate{1}{U (\vec{\theta})} & \qswap \qwx & \multigate{1}{U (\vec{\theta})} & \qw \\
\lstick{ {q}_{7} :  }  & \qw & \ghost{U (\vec{\theta})} & \qw & \qw & \qw & \ghost{U (\vec{\theta})} & \qswap & \ghost{U (\vec{\theta})}& \qswap  \\
\vdots  &  & \vdots & \vdots & & & \vdots & \qwx & \vdots & \qwx \\
\lstick{ {q}_{n-4} :  } & \qw & \multigate{1}{U (\vec{\theta})} & \qw & \qw & \qw & \multigate{1}{U (\vec{\theta})} & \qswap \qwx & \multigate{1}{U (\vec{\theta})} & \qswap \qwx \\
\lstick{ {q}_{n-3} :  } & \qw & \ghost{U (\vec{\theta})} & \multigate{1}{U (\vec{\theta})} & \qw & \qswap & \ghost{U (\vec{\theta})} & \qswap & \ghost{U (\vec{\theta})}& \qw \\
\lstick{ {q}_{n-2} :  } & \qw & \multigate{1}{U (\vec{\theta})} & \ghost{U (\vec{\theta})} & \qw & \qswap \qwx & \multigate{1}{U (\vec{\theta})} & \qswap \qwx & \multigate{1}{U (\vec{\theta})} & \qw  \\
\lstick{ {q}_{n-1} :  } & \qw & \ghost{U (\vec{\theta})} & \qw & \qw & \qw & \ghost{U (\vec{\theta})} & \qw & \ghost{U (\vec{\theta})}& \qw
\gategroup{1}{3}{13}{5}{.7em}{-}
\gategroup{1}{6}{13}{10}{.7em}{--}
}
\]
\vspace*{1mm}
\caption{One Trotter step implementation for the Hamiltonian $H_{\mathrm{Dimer}}$. For a periodic boundary condition, the odd layers in the straight line box have the two-qubit gates, $U(\vec{\theta})$, between $q_{n-1}$ and $q_0$ and the swap gate between $q_{n-1}$ and $q_0$ in the second and the third swap gate layers in the dotted line box.}
\label{fig:circuit_HE}
\end{figure}
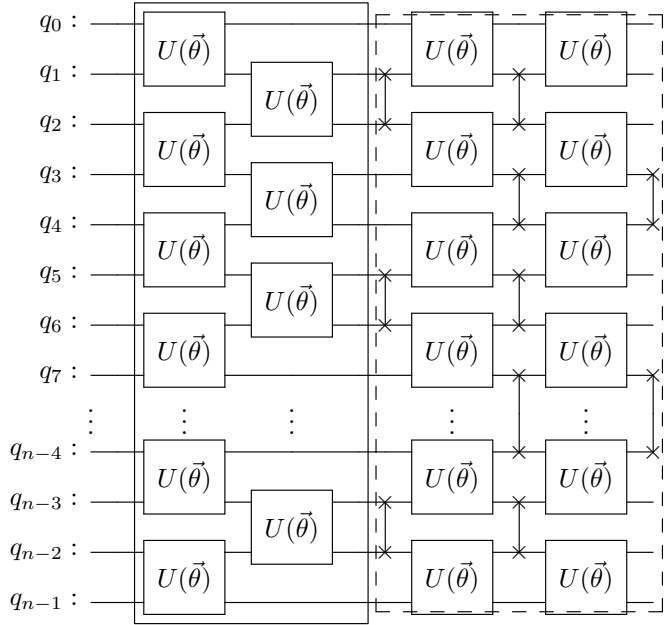 

The Dimer Hamiltonian $H_{\mathrm{Dimer}}$ has additional terms in addition to the Heisenberg XXX spin chain Hamiltonian $H_{\mathrm{iso}}$  as shown in Eq.~\eqref{eq:hamiltonian}. The additional terms have coefficient $J_2$. The $J_2$ terms have interaction with the next nearest neighbor sites. 
Hence, the Dimer Hamiltonian has the interaction with the nearest neighbor and the next nearest neighbor sites.
This is the main challenge to make a quantum circuit for the time evolution of the Hamiltonian on quantum computers having limited connectivity between qubits such as in superconducting quantum computers. 

In this section, we describe our new quantum circuit design for the Dimer Hamiltonian on the quantum devices having only connection between nearest neighbor qubits. That is, all qubits have two connections except the first and the last qubits (a path graph) on open boundary conditions. In periodic boundary conditions, the first and the last qubits are connected (a circle graph).
Figure.~\ref{fig:circuit_HE} shows one Trotter step of the first-order Trotterization for the Dimer Hamiltonian. In the figure, the box surrounded by the straight line is the isotropic Heisenberg Hamiltonian part (cf. Fig.~\ref{fig:circuit_XXX_1st}).
The dotted box part in Fig. \ref{fig:circuit_HE} represents our new circuit design for the $J_2$ terms.
In this circuit design, we assume that the $n$ is a multiple of $4$.
In the numbering notation, the upper bar of a number represents a modulo number of $4$. For example, $\bar{k}$ means $k$ mod $4$.
In addition, $\text{SWAP}(i, j)$ represents the swap gate between $q_i$ and $q_j$.
The first step is placing $\text{SWAP}(\bar{1}, \bar{2})$. This is depicted in the first swap layer in the dotted box in Fig. \ref{fig:circuit_HE}. 
The second is placing the even layer on the whole qubits.
The first swap gate layer and the following even layers process the $J_2$ terms between $\bar{0}$ and $\bar{2}$, and $\bar{1}$ and $\bar{3}$, respectively.
The third is placing $\text{SWAP}(\bar{1}, \bar{2})$, and $\text{SWAP}(\bar{3}, \bar{0})$ of the neighbor. If it is under PBC, $\text{SWAP}(n-1, 0)$ is added.
$\text{SWAP}(\bar{1}, \bar{2})$ is reversing the first step, $\text{SWAP}(\bar{1}, \bar{2})$.
The fourth is placing the even layer on the whole qubits.
$\text{SWAP}(\bar{3}, \bar{0})$ of the neighbor and the following even layers process the $J_2$ terms between $\bar{2}$ and $\bar{0}$, and $\bar{3}$ and $\bar{1}$, respectively.
Finally, $\text{SWAP}(\bar{3}, \bar{0})$ of neighbor are placed. If it is under PBC, $\text{SWAP}(n-1, 0)$ is added. 
The final process is reversing the $\text{SWAP}(\bar{3}, \bar{0})$ of the neighbor in the third process.
The whole process of the circuit construction is summarized in TABLE~\ref{tab:alg}.

\begin{table}[h!]
\begin{center}
\begin{tabular}{rl}

\hline \hline
    \multicolumn{2}{l}{\textbf{Notations}:}\\
    \multicolumn{2}{l}{$n$ is the number of qubits and a multiple of $4$.}\\
    \multicolumn{2}{l}{The numbers represent qubit index from $0$ to $n-1$.}\\
    \multicolumn{2}{l}{$\bar{k}$ represent $k$ modulo $4$.}\\
\hline

1: & Place swap gates between $\bar{1}$ and $\bar{2}$.\\ 
2: & Place the even layer of $\vec{\theta}$ on the whole qubits.\\
3: & Place swap gates between $\bar{1}$ and $\bar{2}$, and between $\bar{3}$ and $\bar{0}.$\\
  & if PBC, place swap gate between $n-1$ and $0$.\\
4: & Place the even layer of $\vec{\theta}$ on the whole qubits. \\
5: & Place swap gates between $\bar{3}$ and $\bar{0}$.\\
  & if PBC, place swap gate between $n-1$ and $0$.\\
\hline
\hline
\end{tabular}
\caption{Summary of the circuit construction for the $J_2$ terms in Fig. \ref{fig:circuit_HE}.}
\label{tab:alg}
\end{center}
\end{table}

\section{Implementation for Experiments}\label{sec:experimental}

As we discussed in the previous section, the basic building block for the quantum circuit implementation is $U_i (\vec{\theta})$ in Eq.~\eqref{eq:evolution_XYZ} for both the isotropic Heisenberg Hamiltonian $H_{\mathrm{iso}}$ and the Dimer Hamiltonian $H_{\mathrm{Dimer}}$. Hence, the key to a successful simulation on noisy quantum computers lies in the implementation of an efficient quantum circuit, as contemporary noisy quantum computers are susceptible to various quantum noise sources, including quantum gate errors. In this section, we describe our specific circuit implementation of Eq.~\eqref{eq:evolution_XYZ} to execute the time evolution on the IBM quantum computers.
Sec.~\ref{sec:quantum_circuit} summarizes the quantum circuit implementation. 
As described in Sec.~\ref{sec:time-evoliton}, we need only an efficient implementation of Eq.~\eqref{eq:evolution_XYZ} for the time evolution of the isotropic Heisenberg Hamiltonian ($H_{\mathrm{iso}}$) and the Dimer Hamiltonian ($H_{\mathrm{Dimer}}$) as well as swap gates between the nearest neighbor qubits.
Since our new circuit implementations for $H_{\mathrm{iso}}$ and $H_{\mathrm{Dimer}}$ only use quantum gates working on the nearest neighbor qubits,  these implementations circumvent the limited qubit connection issue of IBM quantum computers.
Based on the quantum circuit implementation, various quantum error mitigation methods are applied and the methods are described in Sec. \ref{sec:error_mitigations}.

\subsection{Quantum Circuit Implementation}\label{sec:quantum_circuit}

To implement Eq.~\eqref{eq:evolution_XYZ}, we start from the Ising coupling gate, $R_{Z_i Z_j} (\theta)$ as follows:
\begin{align}
\nonumber
R_{Z_i Z_j}(\theta) = \exp\left(-i \frac{\theta}{2} \sigma_i^z \sigma_j^z \right) = \begin{pmatrix} e^{-i \frac{\theta}{2}} & 0 & 0 & 0 \\ 0 & e^{i \frac{\theta}{2}} & 0 & 0 \\ 0 & 0 & e^{i \frac{\theta}{2}} & 0 \\ 0 & 0 & 0 & e^{-i \frac{\theta}{2}}  \end{pmatrix}
\end{align}
which is implemented as \texttt{RZZGate} in IBM Qiskit.
Since we have Clifford gate identities, we have the induced $R_{X_i X_j} (\theta)$ and $R_{Y_i Y_j} (\theta)$ gates as follows:

\begin{align}
\nonumber
\label{eq:Rz_two_cnot}
    \Qcircuit @C=0.7em @R=1.5em {
    &&&&& \gate{H} & \ctrl{1} & \qw & \ctrl{1} & \gate{H} & \\
    \raisebox{1.2cm}{$R_{XX}(\theta)=$} &&&&& \gate{H} & \targ & \gate{R_z(\theta)} & \targ & \gate{H} & , \\
    &&&&& \gate{\sqrt{\sigma^x}} & \ctrl{1} & \qw & \ctrl{1} & \gate{\sqrt{\sigma^x}^{\dagger}} & \\
    \raisebox{1.2cm}{$R_{YY}(\theta)=$} &&&&& \gate{\sqrt{\sigma^x}} & \targ & \gate{R_z(\theta)} & \targ & \gate{\sqrt{\sigma^x}^{\dagger}} & , \\
    &&&&&& \ctrl{1} & \qw & \ctrl{1} & \qw & \\
    \raisebox{0.8cm}{$R_{ZZ}(\theta)=$} &&&&&& \targ & \gate{R_z(\theta)} & \targ & \qw & 
    }
\end{align}
where $H$ is the Hadarmard gate, $\sqrt{\sigma^x} = \frac{1}{2} {\begin{pmatrix} 1+i & 1-i \\ 1-i & 1+i \end{pmatrix}}$, and $R_z (\theta) = {\begin{pmatrix} e^{-i \frac{\theta}{2}} & 0 \\ 0 & e^{i \frac{\theta}{2}} \end{pmatrix}}$.
Hence, $U_i (\vec{\theta})$ in Eq.~\eqref{eq:hamiltonian_Pauli} is implemented as follows:
\begin{align}
\nonumber
    \Qcircuit @C=0.7em @R=1.5em {
    &&& \gate{H} & \ctrl{1} & \qw & \ctrl{1} & \gate{H} & \qw & \gate{\sqrt{\sigma^x}} & \qw & \ctrl{1} & \qw & \ctrl{1} & \gate{\sqrt{\sigma^x}^{\dagger}} & \ctrl{1} & \qw & \ctrl{1} & \qw & \\
    &&& \gate{H} & \targ & \gate{R_z(\theta)} & \targ & \gate{H} & \qw & \gate{\sqrt{\sigma^x}} & \qw & \targ & \gate{R_z(\theta)} & \targ & \gate{\sqrt{\sigma^x}^{\dagger}} & \targ & \gate{R_z(\theta)} & \targ & \qw & 
    }
\end{align}
and this implementation has six \texttt{CX} gates and thirteen circuit depths.
This circuit is compressed and optimized by circuit identities as follows:

\begin{equation}
\label{eq:circuit_XYZ}
\Qcircuit @C=1.0em @R=0.7em @!R { 
 & \targ & \gate{R_Z ( \theta_z )} & \qw & \targ & \gate{R_Z (- \theta_y)} & \targ & \gate{\sqrt{\sigma^x}} & \qw & \qw & \\
 & \ctrl{-1} & \gate{H} & \gate{R_Z (\theta_x + \frac{\pi}{2})} & \ctrl{-1} & \gate{H} & \ctrl{-1} &  \gate{\sqrt{\sigma^x}^\dagger} & \qw & \qw  
}
\end{equation}
and this circuit has three \texttt{CX} gates and seven depths.
The induction of the circuit identity is described in detail in Appendix A in Ref. \cite{zhang2024optimal}.
Since quantum gates have gate errors and two-qubit gates such as \texttt{CX} are noisier than single-qubit gates, reducing the number of \texttt{CX} gates as well as the circuit depths is essential to reduce the overall noise on the quantum computers. 
Hence, we adopt the quantum circuit described in Eq.~\eqref{eq:circuit_XYZ} for the implementation of Eq.~\eqref{eq:evolution_XYZ}.

\subsection{Quantum Error Mitigations}\label{sec:error_mitigations}

The troublesome challenge of running quantum algorithms on contemporary quantum devices, including IBM Quantum processors, is the errors and noise on the quantum devices
To cope with the errors and noises, quantum error correction (QEC) was suggested \cite{shor1995scheme, calderbank1996good}.
However, QEC has a qubit overhead that is daunting to implement on a large problem on the contemporary quantum processors even though they are optimized \cite{kivlichan2020improved, lee2021even}.
On the other hand, quantum error mitigation (QEM) accepts the imperfection of contemporary quantum devices and adopts methods of mitigating or suppressing quantum errors and noises.
QEM has a low or no qubit overhead. 
In recent years, various QEM methods have been developed, and their practicality has been proven in practical problems \cite{yu2023simulating, Kim-error-mitigation, kim2023evidence, charles2305simulating}. 
Hence, we apply four QEM methods, Zero-Noise Extrapolation (ZNE), Pauli Twirling (PT), Dynamical Decoupling (DD), and Matrix-free Measurement Mitigation (M3) to cope with the quantum device errors and noises in our experiments.
A quantum circuit for one Hamiltonian simulation at a time is extended to three variational circuits including itself for the ZNE method and each variational circuit is duplicated to ten copies to apply the PT. Hence, we have $30$ circuits for one Hamiltonian simulation at a time.
We explain the QEM methods in detail in the following sub-sections.

\subsubsection{Zero-Noise Extrapolation}

Zero-noise extrapolation (ZNE) is a quantum error mitigation method that estimates an ideal expectation value (no noise expectation value) from other expectation values at different noise levels by extrapolation methods \cite{Temme-error-mitigation, Li-error-mitigation, giurgica2020digital}.
In our experiments, we adopted local unitary gate folding \cite{giurgica2020digital} with scaling factors $1, 3,$ and $5$ only on two-qubit gates such as $\texttt{CX}$ (or $\texttt{CZ}$. Refer to Sec. \ref{sec:circuit_construction}) gates rather than applying the folding to all gates since the two-qubit gate is more than ten times noisier than single qubit gates.

\subsubsection{Pauli Twirling}

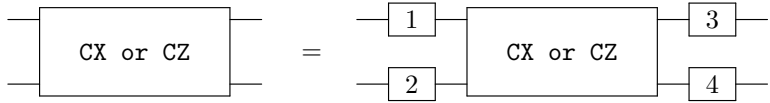
\begin{figure}[h!]
\[
\Qcircuit @C=1.2em @R=1.2em {
  & \multigate{1}{~~~\texttt{CX or CZ}~~~} & \qw & & & & \gate{~1~} & \multigate{1}{~~~\texttt{CX or CZ}~~~} & \gate{~3~} & \qw \\
  & \ghost{~~~\texttt{CX or CZ}~~~} & \qw &  & \raisebox{0.8cm}{~~=~~~~~} & & \gate{~2~} & \ghost{~~~\texttt{CX or CZ}~~~} & \gate{~4~}  & \qw
}
\]
\vspace*{1mm}
\caption{A circuit diagram for Pauli twirling for $\texttt{CX}$, $\texttt{CZ}$, or $\texttt{ECR}$ gates. At positions 1, 2, 3, and 4, Pauli gates, $\{ I, \sigma^x, \sigma^y, \sigma^z \}$ are placed.} 
\label{fig:pauli_twriling}
\end{figure}

Pauli twirling is a method averaging out the off-diagonal coherent errors of the circuits in the Pauli basis, $\{ I, \sigma^x, \sigma^y, \sigma^z \}$ \cite{bennett1996purification, wallman2016noise, cai2019constructing}.
In Pauli twirling, a Clifford gate is surrounded by the Pauli gates back and forth which is mathematically identical to the Clifford gate.
The efficiency is empirically proved in previous studies \cite{Kim-error-mitigation, kim2023evidence}.
Figure~\ref{fig:pauli_twriling} shows the Pauli twirling method we applied to this study.
First, we searched all the combinations of Pauli gates that are mathematically identical up the global phase with only the Clifford gate ($\texttt{CX}$ or $\texttt{CZ}$ gate depending on the target quantum device in our experiments).
This is our Pauli twirling gates set.
Since the Pauli gate set has four elements and we have four positions, the search domain is $256 (=4^4)$ cases.
The trivial case is placing the identity gate in the position, $1, 2, 3$, and $4$ in Fig. \ref{fig:pauli_twriling}.
One non-trivial case is putting $\sigma^z, \sigma^x, \sigma^z,$ and $\sigma^x$ at $1, 2, 3$, and $4$, respectively.
We duplicated $10$ copies of the base quantum circuit.
After that, we randomly chose a Pauli twirling gate combination out of the prepared Pauli twirling gates set and applied them to the Clifford gate as described in Fig. \ref{fig:pauli_twriling}.

\subsubsection{Dynamical Decoupling}

Dynamical decoupling (DD) is a quantum error mitigation method that reduces errors caused by spectator qubits.
DD is implemented by periodic sequences of instantaneous control pulses that average out the coupling with the system environment to approximately zero \cite{viola1999dynamical}.
In particular, a set of single qubit operators are interleaved using basis transformation on idle qubits so that environmental contamination from other qubits is decoupled. 
Consequently, the coherence time of the circuit becomes longer.
The efficiency of DD is empirically tested in various environments \cite{ezzell2022dynamical, niu2022effects, Kim-error-mitigation, kim2023evidence, charles2305simulating}.
In this study, we added ($t/4$, $\sigma^x$, $t/2$, $\sigma^x$, $t/4$) sequence in every idling period through Qiskit $\texttt{PassManager}$ where $t$ is the idling time except for the two $\texttt{XGate}$ pulse durations.

\subsubsection{Measurement Error Mitigation}

The canonical measurement error mitigation methods \cite{bravyi2021mitigating, geller2020rigorous, hamilton2006scalable} correct the measurement error over $N$ qubits by computing the measurement error probability matrix as follows:
\begin{align}
\nonumber
 \vec{s}_{\textrm{noisy}} = \mathcal{M} \vec{s}_{\textrm{ideal}}
\end{align}
where $\vec{s}_{\textrm{noisy}}$ and $\vec{s}_{\textrm{ideal}}$ are a state vector of noisy probabilities returned by the quantum system, and a state vector of the probabilities in the absence of measurement errors, respectively. 
Since $\vec{s}_{\textrm{noisy}}$ and $\vec{s}_{\textrm{ideal}}$ are state vector of $N$ qubit, the matrix $\mathcal{M}$ has $2^N \times 2^N$ dimension with entry $A_{i, j}$ is the probability of bit string $j$ being converted to bit string $i$ by the measurement-error process.
Although errors across multiple qubits can be accurately approximated by employing no more than $\mathcal{O} (N)$ calibration circuits, 
the method has to compute the inverse of $\mathcal{M}$ in order to estimate the ideal measurement after getting noisy measurement results.
This makes the method impractical at large qubit numbers.
Instead of the canonical measurement error mitigation methods, a matrix-free measurement mitigation (M3) method has been invented \cite{nation2021scalable}.
The method, M3, works in a reduced subspace determined by the noisy input bitstrings requiring correction. This space often contains significantly fewer unique bitstrings compared to the expansive multi-qubit Hilbert space, making the resulting set of linear equations notably simpler to resolve.
This method is implemented in Python \cite{mthree}.
Since we conducted the experiments with $20, 96,$ and $100$ qubits, it was not possible to use the canonical measurement error mitigation methods. So, We adopted M3 for our measurement error mitigation by using the implementation \cite{mthree}.

\subsection{Circuit Implementation for Experiments and Post Processing}
\label{sec:circuit_construction}

In this study, we used two 127-qubit IBM quantum processors, $\texttt{ibm\_sherbrooke}$, $\texttt{ibm\_brisbane}$, and one 133-qubit IBM quantum processor, $\texttt{ibm\_torino}$.
The 127-qubit processors are IBM Eagle r3 quantum processors and they have a basis gate set, $\{ \texttt{ECR}, \texttt{I}, \texttt{RZ}, \texttt{SX}, \texttt{X} \}$ where $\texttt{I}$, $\texttt{RZ}$, $\texttt{SX}$, and $\texttt{ECR}$ are the identity, $R_z$, $\sqrt{\sigma^x}$, and $\frac{1}{\sqrt{2}} \left( IX - XY \right)$, respectively \cite{chow2011simple, corcoles2015demonstration, IBMQ-ECR}.
On the other hand, the 133-qubit processors adopt the IBM Heron r1 processor type. This type has the basis gate set, $\{ \texttt{CZ}, \texttt{I}, \texttt{RZ}, \texttt{SX}, \texttt{X} \}$ where $\texttt{CZ}$ is a controlled-$\texttt{Z}$ gate, $I \otimes \ket{0}\bra{0} + Z \otimes  \ket{0}\bra{0}$ or $Z \otimes \ket{0}\bra{0} + I \otimes  \ket{0}\bra{0}$ (Refer to Sec. \ref{sec:quantum_circuit} for the quantum gate definitions).
In the basis sets, $\texttt{ECR}$ and $\texttt{CZ}$ are two-qubit gates and other gates are one-qubit gates.
These two-qubit gates are primitives for constructing $\texttt{CX}$ gate.
In our circuit construction, we used $\texttt{CX}$ gates regardless of the target devices. The Qiskit transpiler converts all other gates into the gates in the basis gate set depending on the target devices.

To construct the quantum circuit with the error mitigation methods described in Sec. \ref{sec:error_mitigations}, we first transpiled the quantum circuits of each Trotter step.
The circuit implementation of the Trotter steps is described in Sec. \ref{sec:time-evoliton} and Sec. \ref{sec:quantum_circuit}.
For the transpiling, we used the qubit mapping visualized in Fig. \ref{fig:QMapping}. 
During the transpiling process, the logical qubits are mapped to physical qubits as described in Fig. \ref{fig:QMapping}, and the logical quantum gates are converted into sets in the basis gate set with circuit optimizations.
We applied the highest optimization level during the transpiling.

After that, we duplicated each circuit two times (three circuits including the base circuit) and applied a local unitary gate folding with scaling factors 1, 3, and 5 to the three circuits, respectively.
In the next place, we applied the Pauli twirling.
We duplicated each circuit $10$ times including the base circuit and surrounded the Clifford gate with a randomly chosen Pauli twirling combination out of the prepared Pauli twirling gates set as described in Fig. \ref{fig:pauli_twriling}.
Finally, we applied the dynamical decoupling method to all those circuits.
Up to now, we have thirty circuits for a Trotter step.

To execute each circuit, we used $10,000$ shots (repeated circuit execution for the measurement sampling) in all cases.
At the end of the circuit executions, we applied the measurement error mitigation method. 
We used a Python library \cite{mthree} to calibrate the library from the system error information and to correct the measurement errors.
After gathering all the measurement results of $10$ circuit duplication for the Pauli twirling, the expectation values are computed for each ZNE folding copy (1, 3, and 5 scaling factors).
Finally, the ZNE is estimated by a quadratic polynomial fitting curve.

\section{Results and Discussion}\label{sec:results}
In order to ensure the accuracy of the results obtained from quantum computers, it is crucial to cross-check those results with classical numerical methods used to study the many-body system. Nevertheless, the classical approach becomes inefficient with the number of qubits, given the exponential growth in the dimensionality of the Hilbert space and therefore importance of quantum computing becomes crucial for large-scale calculations. In the following, we briefly present our two adopted classical methods for checking the measured values from IBM's quantum devices.

\noindent\textbf{Direct method:} One straightforward approach, denoted as the \textit{direct} method in this work, is to calculate the time-evolved expectation value of the staggered magnetization $\hat{O}_{M_{st}}$ for $N$ qubits with respect to the N\'eel state, $\langle\psi_{\mathrm{Neel}}(t)|\hat{O}_{M_{st}}|\psi_{\mathrm{Neel}}(t)\rangle$ where $|\psi_{\mathrm{Neel}}(t)\rangle=e^{-i\,H\,t}|\psi_{\mathrm{Neel}}\rangle$. The Hamiltonian $H$ is either the isotropic Hamiltonian $H_{\mathrm{iso}}$ or the Dimer Hamiltonian $H_{\mathrm{Dimer}}$ of our study. Here, $H$ is a $2^{N}\times 2^{N}$ Hermitian matrix that acts on the Hilbert space of dimension $2^{N}$. We have implemented this approach using our own Python implementation and have checked the results with QuSpin~\cite{quspin}. This method is the simplest and most accurate. However, it becomes an inefficient computational mode for calculating the time evolution of state vectors with $N\simgt 20$ qubits, as the Hilbert space's dimensionality increases exponentially with the number of qubits. For example, the number of qubits $N=50$ requires 16 petabytes of memory allocation in double precision for expressing just a state vector, which is possible only for present supercomputers. Therefore, we turn to the classical approximation method based on matrix product states to calculate the time evolution of state vectors with $N\simgt 20$ qubits.

\noindent\textbf{MPS-TDVP method:} Matrix product states (MPS) is a common method used to study the time evolution of large quantum many-body systems~\cite{Paeckel:2019yjf, Cirac:2020obd}. MPS is a one-dimensional array of tensors linked together, with each tensor corresponding to a site or particle of the many-body system. The indices connecting the tensors in the MPS are called bond indices, which can take up to $\chi$ values (also known as bond dimensions). Meanwhile, the open indices of each tensor correspond to the physical degrees of freedom of the local Hilbert space associated with a site or a particle of the system, which can take up to $d$ values (for our system of spin-1/2 particles, $d=2$). While the MPS can represent any quantum state of the many-body system, the bond dimension $\chi$ needs to be exponentially large in the system size to cover all states in the Hilbert space. We determine the time evolution of the expectation value of staggered magnetization for the N\'eel state, which we denote here as the \textit{MPS-TDVP} method, using approximation method based on time-dependent variational principle (TDVP)~\cite{TDVP-ref-1, TDVP-ref-2}, facilitated by the package ITensor~\cite{itensor-1, itensor-2, itensor-TDVP}. The time evolution of MPS using the TDVP-based method is advantageous as it can not only handle the Hamiltonian with long-range interactions rather than only the nearest-neighbor interaction but is also computationally less demanding when the PBC is imposed on those Hamiltonians.

In the subsequent analysis, we consider the time parameter $t$ in an arbitrary unit where $\hbar=1$ and $J_{1}=1$. One can simply restore $t$ in seconds by mapping $t\rightarrow \hbar t/J_{1}$. For a typical value of exchange interaction, $J_{1}\sim O(\mathrm{eV})$ that is associated with magnetic materials, $t$ falls in $O(10^{-15})$ sec, the time-scale of atomic transitions. Besides, we choose $\delta t=0.1$ as the time-step size and maximum allowed error $\epsilon=10^{-12}$ for each sweep in the MPS-TDVP method, which leads to 10 and 40 sweeps when we evolve the system up to $t=1$ and $t=4$, respectively.

\subsection{$N=20$ Qubits}
\label{sec:n20}

\begin{figure}[h!]
    \centerline{
    \includegraphics[width=0.55\textwidth]{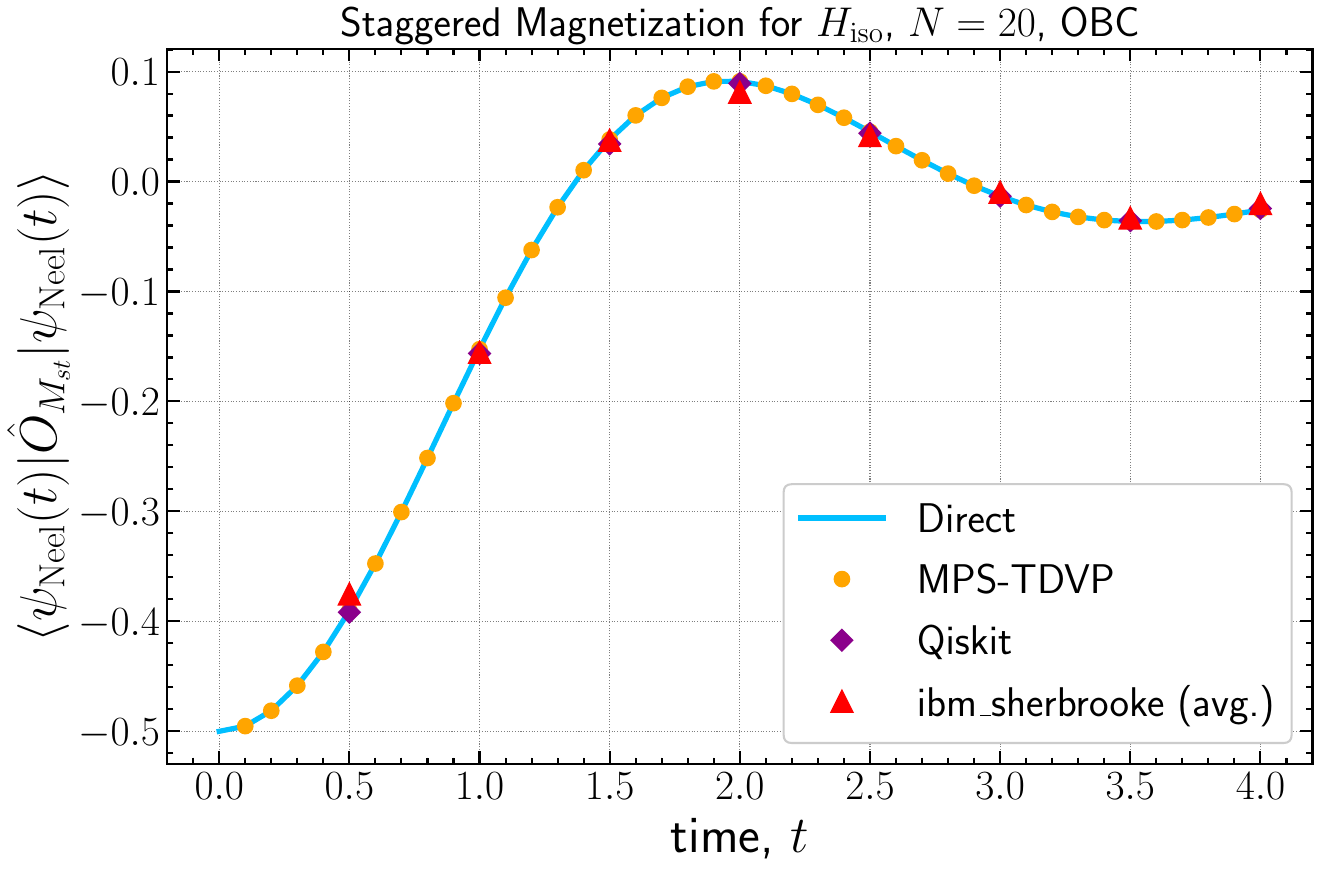}\hspace{1mm}
    \includegraphics[width=0.55\textwidth]{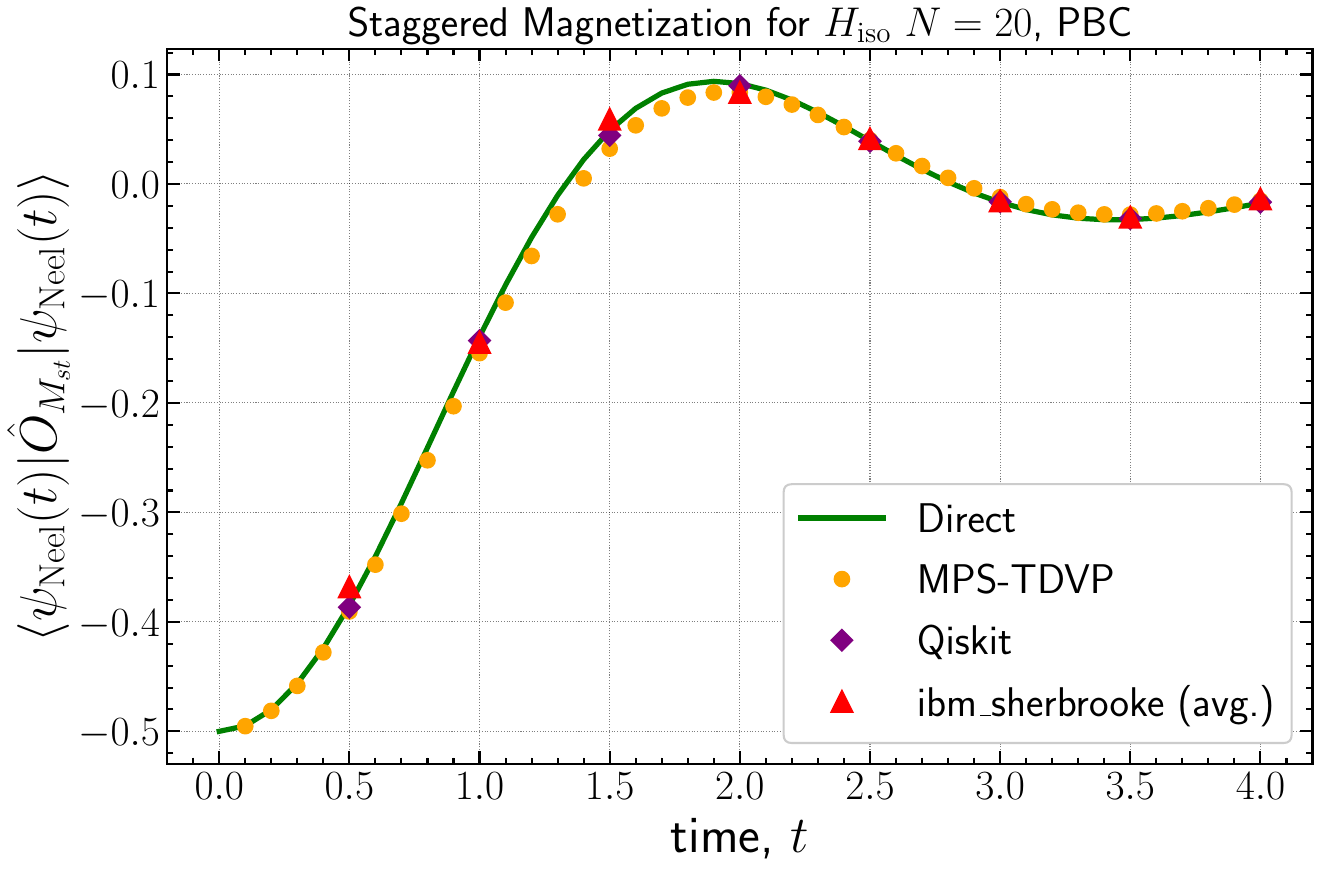}}
    \caption{Time evolution of the expectation value of staggered magnetization for the N\'eel state under the Hamiltonian $H_{\mathrm{iso}}$ for $N=20$ qubits with OBC (left) and PBC (right), respectively.}
    \label{fig:Op_Neel_HB_N=20}
\end{figure}

We present the time evolution of the expectation value of the staggered magnetization observable for the N\'eel state under the isotropic Heisenberg Hamiltonian $H_{\mathrm{iso}}$ using the second-order Trotterization (Sec \ref{sec:time-evoliton_XXX}) in Fig.~\ref{fig:Op_Neel_HB_N=20} for $N=20$ qubits with OBC and PBC cases, respectively.
In the experiments on \texttt{ibm$\_$sherbrooke}, we applied the error mitigation techniques which are already delineated in Sec.~\ref{sec:error_mitigations}, and ran each circuit with $100,000$ shots (trials).
This process was repeated five times.
The Qiskit simulation is a Qiskit sampling simulation (\texttt{qasm\_simulator}) of the circuits with $100,000$ shots. The Qiskit experiment is executed once.

From Fig.~\ref{fig:Op_Neel_HB_N=20} (left), we can see an excellent agreement among the classical computations (both direct and MPS-TDVP methods), the Qiskit simulation, and the \texttt{ibm$\_$sherbrooke} experiments.
The plot of \texttt{ibm$\_$sherbrooke} is the average of the five executions.
We tabulate the average values and the standard deviations in Table \ref{tab:HBN20} in the Appendix \ref{app:staggered}.
In contrast, in Fig.~\ref{fig:Op_Neel_HB_N=20} (right), for PBC, though we also have an excellent agreement among the direct, the Qiskit simulation, and the \texttt{ibm$\_$sherbrooke} experiments, there is a mismatch between the results obtained from the direct and MPS-TDVP methods. This mismatch between the two methods is inherently related to the requirement of a larger bond dimension due to the linking between the tensors at the first and last sites of the many-body system and the resulting larger truncation error compared to the OBC cases. We note that for $H_{\mathrm{iso}}$ and $N=20$ qubits with OBC, the maximum link dimension results in $\chi=102$ after 40 sweeps as we evolve the system up to $t=4$. On the other hand, for the PBC case, after 40 sweeps, we end up with $\chi=991$ while keeping the error within $\epsilon$.
\begin{figure}[h!]
    \centerline{
    \includegraphics[width=0.55\textwidth]{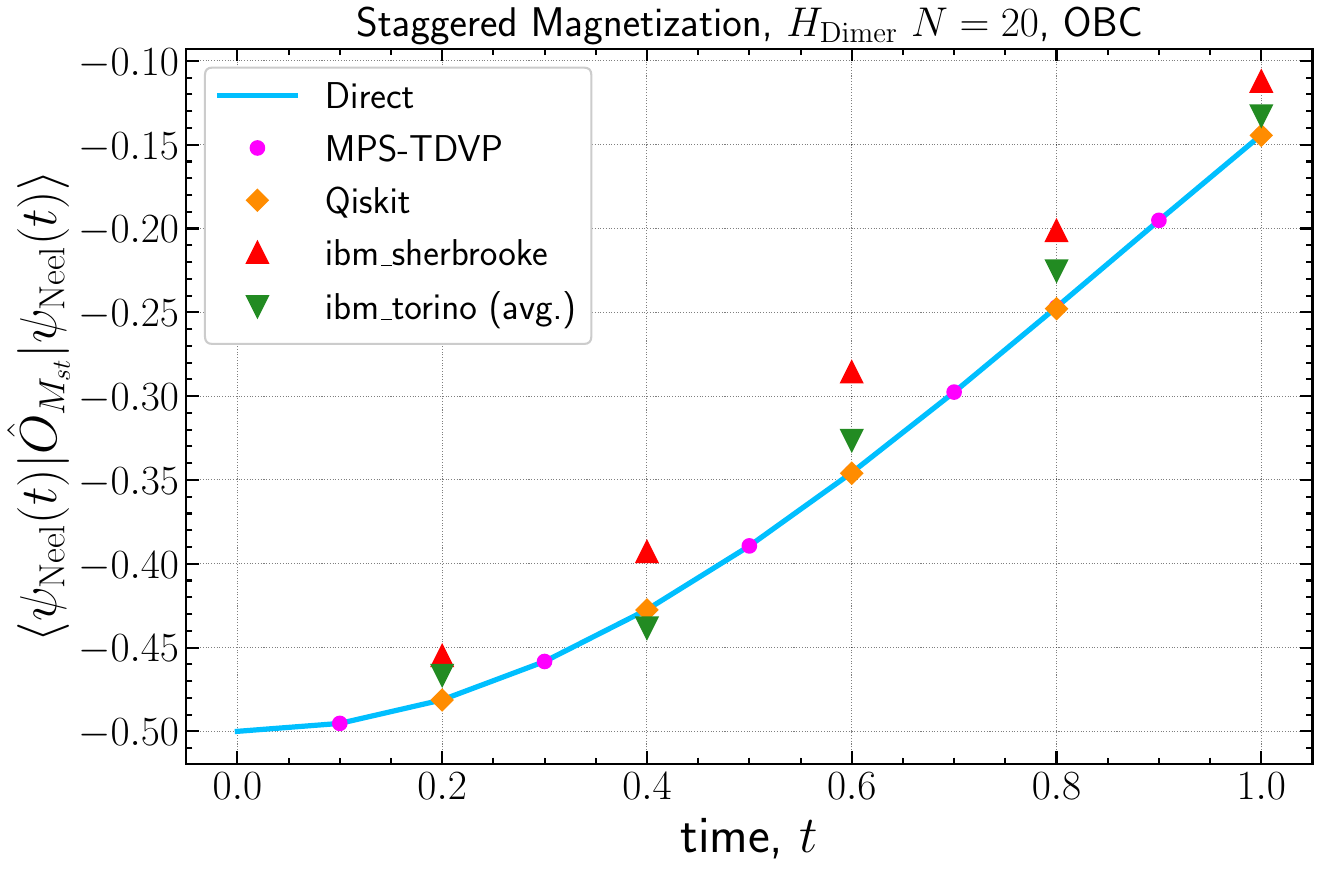}\hspace{1mm}
    \includegraphics[width=0.55\textwidth]{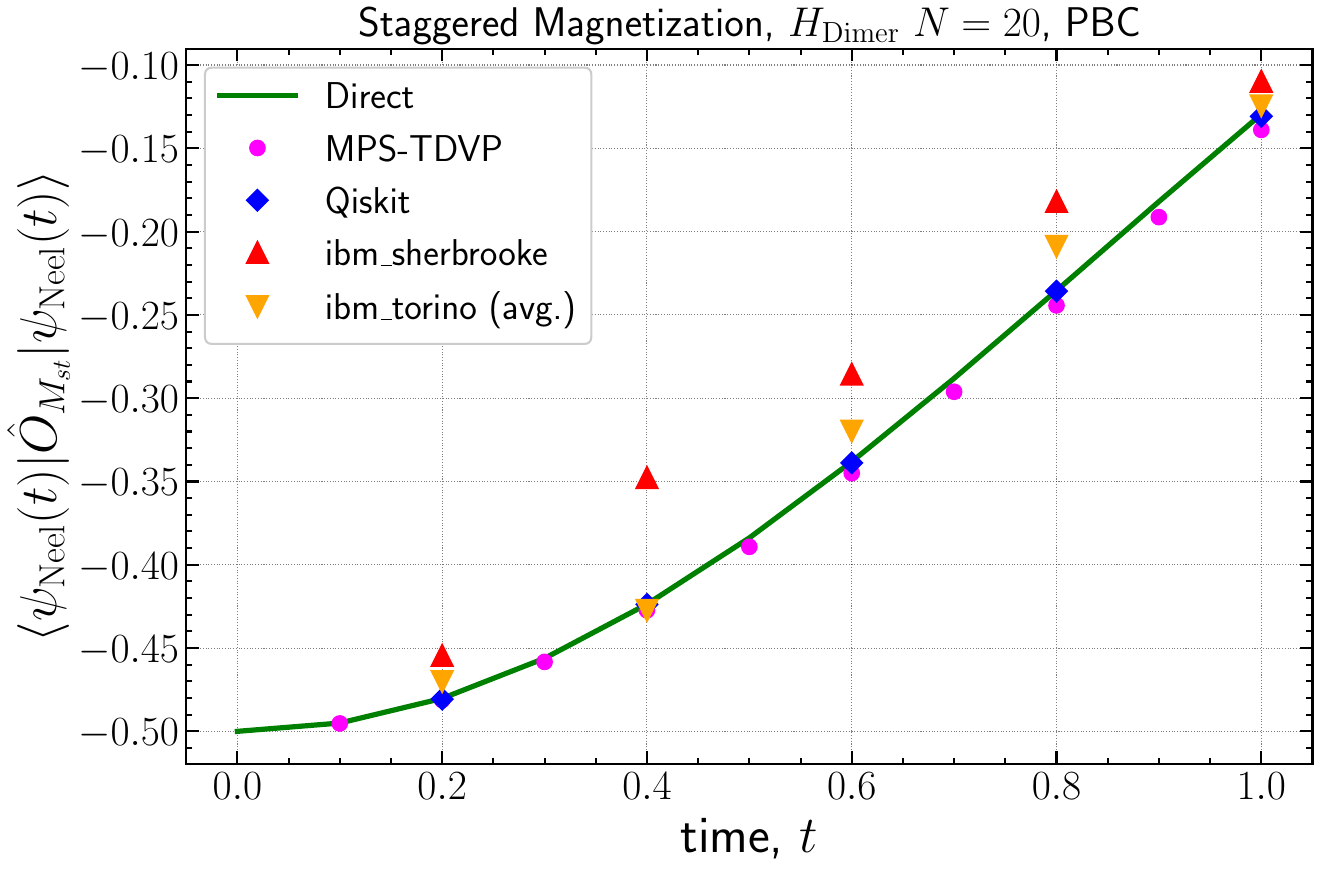}}
    \caption{Time evolution of the expectation value of staggered magnetization with respect to the N\'eel state under the Hamiltonian $H_{\mathrm{Dimer}}$ for $N=20$ qubits.}
    \label{fig:Op_Neel_HE_N=20}
\end{figure}

Figure.~\ref{fig:Op_Neel_HE_N=20} presents the experimental results of $\langle\psi_{\mathrm{Neel}}(t)|\hat{O}_{M_{st}}|\psi_{\mathrm{Neel}}(t)\rangle$ for $H_{\mathrm{Dimer}}$ with $N=20$ qubits using the first-order Trotterization (Sec \ref{sec:time-evoliton_XXX_XXX2}) and comparison with classical computations. 
The Qiskit simulation is a Qiskit sampling simulation (\texttt{qasm\_simulator}) of the circuits with $100,000$ shots.
The experiments of \texttt{ibm$\_$sherbrooke} and \texttt{ibm$\_$torino} uses $100,000$ shots with the QEM (Sec.~\ref{sec:error_mitigations}).
The \texttt{ibm$\_$torino} experiment was repeated five times while the \texttt{ibm$\_$sherbrooke} experiment was executed once.
The average values and the standard deviations of \texttt{ibm$\_$torino} are tabulated in Table \ref{tab:HEN20} in the Appendix \ref{app:staggered}.
The direct computation and the Qiskit simulation show great agreement in both boundary conditions. Again, like the case of $H_{\mathrm{iso}}$, we see a mismatch between the direct computation and the MPS-TDVP for the $H_{\mathrm{Dimer}}$ with PBC in Fig. \ref{fig:Op_Neel_HE_N=20} (right). In the case of OBC, we get the maximum link dimension to be $\chi = 28$ after 10 sweeps to evolve up to $t = 1$, whereas, for the PBC, it was $\chi = 251$ after 10 sweeps while keeping the error within $\epsilon$.
In addition, the \texttt{ibm$\_$sherbrooke} results show a larger discrepancy than the \texttt{ibm$\_$torino} results.
We presume the accuracy difference is originated from the hardware accuracy of $\texttt{ibm\_torino}$ and $\texttt{ibm\_sherbrooke}$ which have $0.8 \%$ and $1.7 \%$ EPLG (Error Per Layered Gate), respectively, in a chain of 100 qubits \cite{ibmq-system}. 
Even though the $\texttt{ibm\_sherbrooke}$ results in Fig. \ref{fig:Op_Neel_HE_N=20} show a good agreement, the results for $H_{\mathrm{Dimer}}$ have notable discrepancy in both boundary conditions.
We conjecture that the reason is that the circuit depth and the number \texttt{CX} gates are different in each Trotter step as shown in Table \ref{tab:circuit-depth-HB}, \ref{tab:circuit-depth-HE}, \ref{tab:cxgates-HB}, and \ref{tab:cxgates-HE} in the Appendix \ref{app:circuitdepth}.
Since the $H_{\mathrm{Dimer}}$ Hamiltonian has the additional term (Refer to Eq. (\ref{eq:hamiltonian_Pauli})) and the implementation for the additional term needs two layers \texttt{SWAP} gates, the $H_{\mathrm{Dimer}}$ Hamiltonian implementation needs about 60 \% more circuit depth than the $H_{\mathrm{iso}}$ Hamiltonian implementation.


\subsection{$N=96$ and $N=100$ qubits}

After cross-checking the measurements of staggered magnetization with the real quantum computers for $N=20$ qubits with classical (direct and MPS-TDVP) methods and Qiskit simulations, we present here our main results of large-scale quantum simulation of the Heisenberg spin chain.
In this extension, we used the same circuit implementation methods, error mitigation methods, and the number of shots (100,000) except for the number of qubits.
We used 100 qubits for the OBC. However, since the PBC needs a connection between the first qubit and the last qubit, we adopted 96 qubits for the PBC. The qubit mapping for these cases is depicted in Fig. \ref{fig:QMapping}.

\begin{figure}[h!]
    \centerline{
    \includegraphics[width=0.55\textwidth]{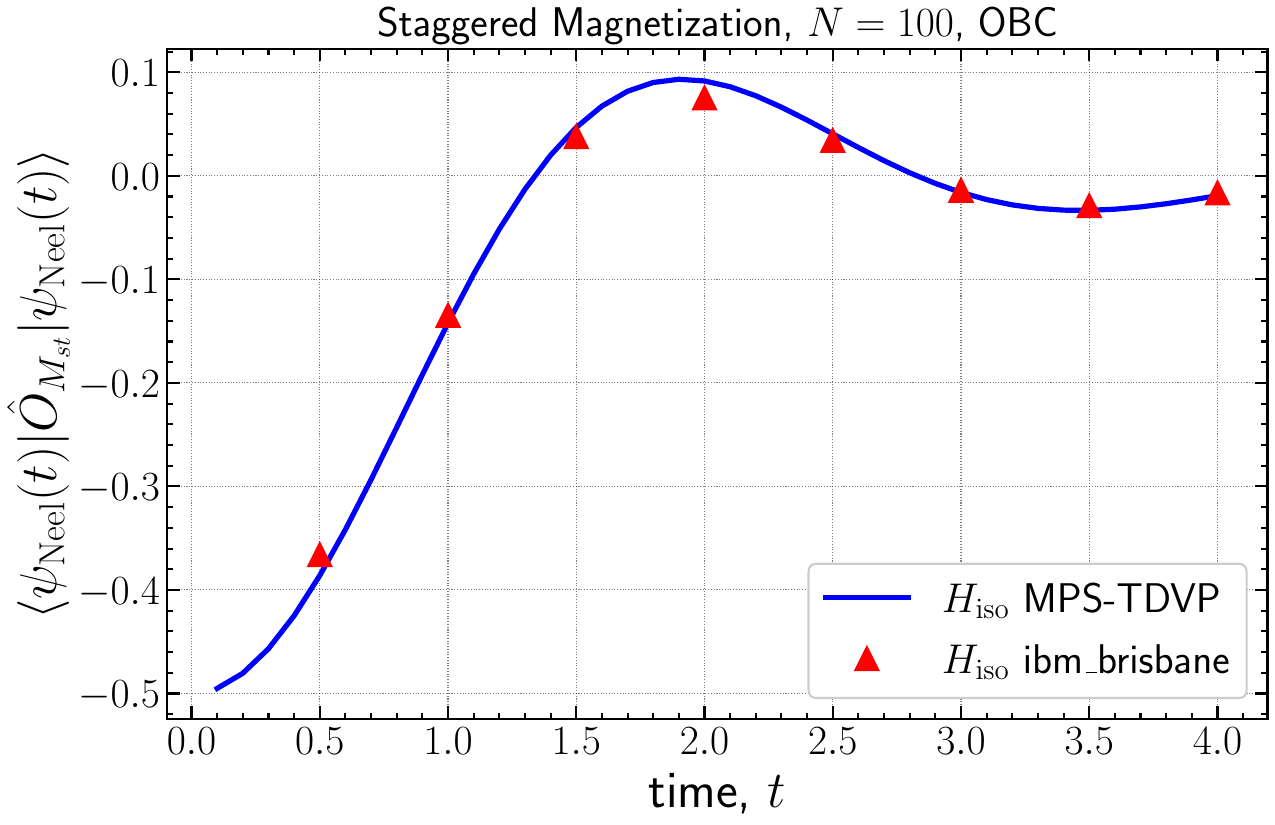}\hspace{1mm}
    \includegraphics[width=0.55\textwidth]{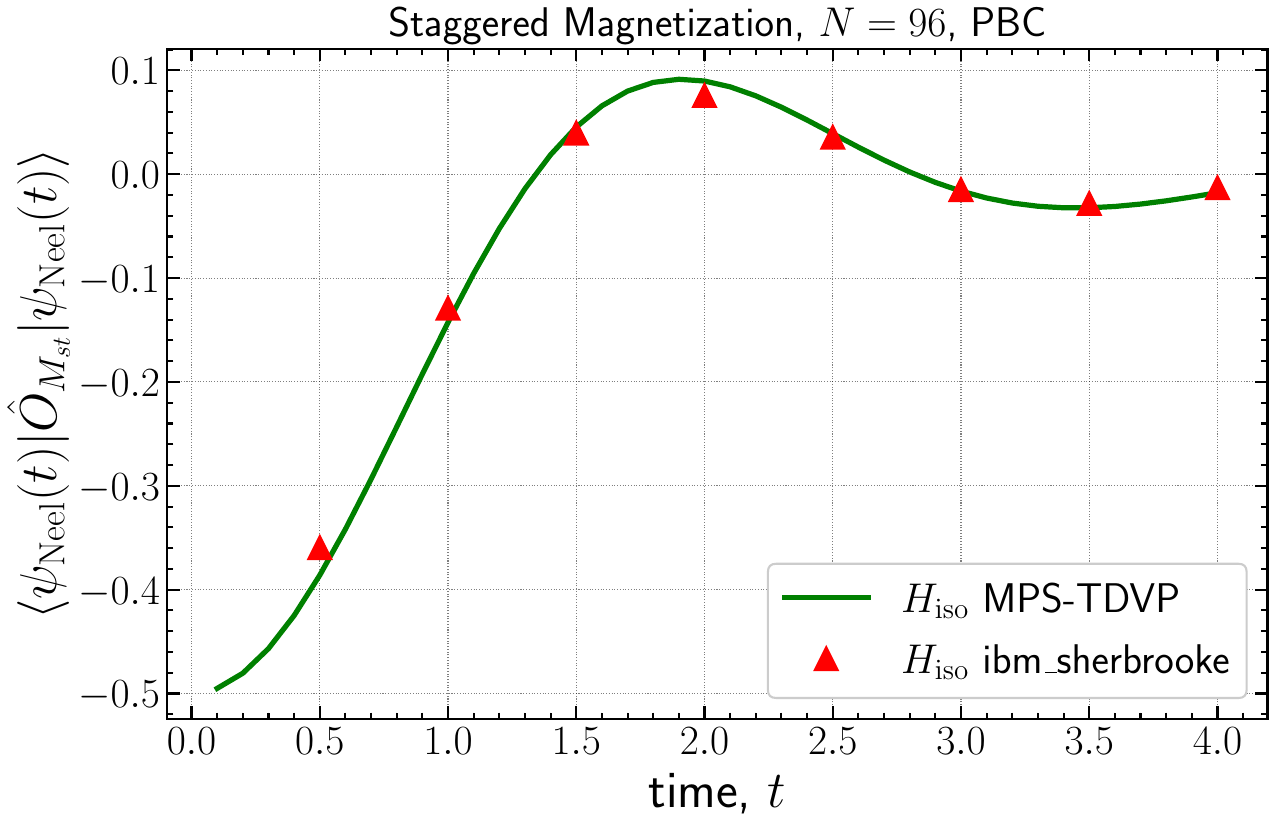}}
    \caption{Time evolution of the expectation value of staggered magnetization for the N\'eel state under the Hamiltonian $H_{\mathrm{iso}}$ for $N=100$ 
 qubits with OBC and $N=96$ qubits with PBC.}
    \label{fig:Op_Neel_HB_large}
\end{figure}

In Fig.~\ref{fig:Op_Neel_HB_large}, we can see that the results of $\langle\psi_{\mathrm{Neel}}(t)|\hat{O}_{M_{st}}|\psi_{\mathrm{Neel}}(t)\rangle$ for $H_{\mathrm{iso}}$ for $N=100$ qubits (OBC) with \texttt{ibm$\_$brisbane} (left figure) and $N=96$ qubits (PBC) with \texttt{ibm$\_$sherbrooke} (right figure), respectively, are in excellent agreement with the results from the MPS-TDVP method. 
Note that for such large-scale systems with $N\sim 100$ qubits, the direct method and the Qiskit simulations are unavailable.

\begin{figure}[h!]
    \centerline{
    \includegraphics[width=0.55\textwidth]{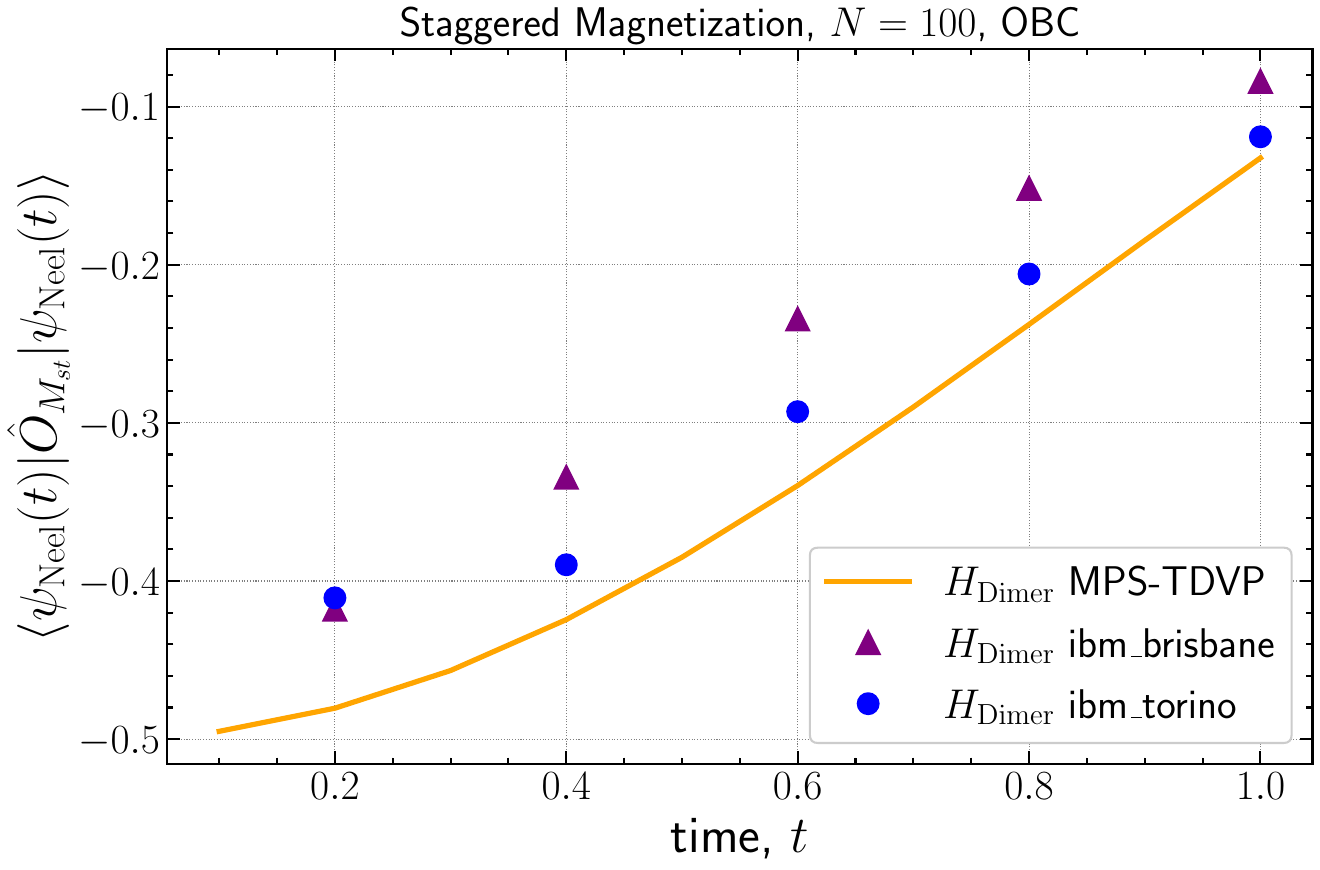}\hspace{1mm}
    \includegraphics[width=0.55\textwidth]{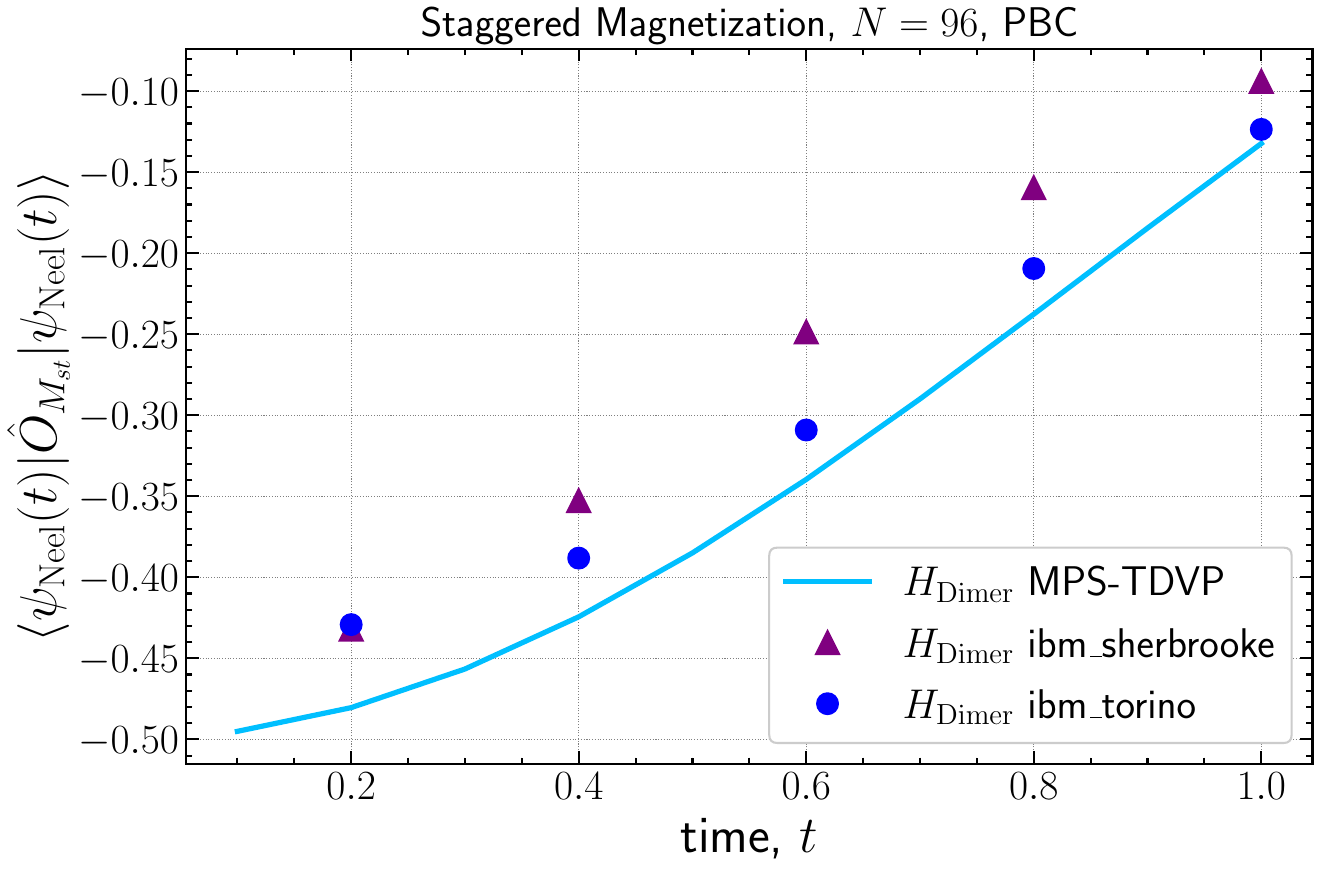}}
    \caption{Time evolution of the expectation value of staggered magnetization for the N\'eel state under the Hamiltonian $H_{\mathrm{Dimer}}$ for $N=100$ qubits with OBC and $N=96$ qubits with PBC.}
    \label{fig:Op_Neel_HE_large}
\end{figure}

Figure~\ref{fig:Op_Neel_HE_large} shows the results of time-evolved staggered magnetization for $H_{\mathrm{Dimer}}$ with time up to $t=1$ with $N=100$ qubits (OBC) (left figure) and $N=96$ qubits (PBC) (right figure) using \texttt{ibm$\_$brisbane}, \texttt{ibm$\_$sherbrooke} and \texttt{ibm$\_$torino}. Likewise the N=20 cases, this case shows a notable discrepancy between the results of the IBM quantum devices and the MPS-TDVP results.

As discussed in Sec. \ref{sec:n20}, we presume that the discrepancy is originated from longer circuit depth and more \texttt{CX} gates in the implementation for the time evolution of $H_{\mathrm{Dimer}}$.
Also, the difference between \texttt{ibm$\_$torino} and \texttt{ibm$\_$brisbane} is derived from the hardware accuracy of $\texttt{ibm\_torino}$ and $\texttt{ibm\_brisbane}$ which have $0.8 \%$ and $1.9 \%$ EPLG (Error Per Layered Gate), respectively, in a chain of 100 qubits \cite{ibmq-system}.

\subsection{Discussion and Future work}
In our experiments, the $H_{\mathrm{Dimer}}$ cases have larger error rates while all the $H_{\mathrm{iso}}$ cases show great accuracy. As
the results from the Qiskit simulation also show better consistency with the direct computation at $N=20$ cases, as seen in Fig.~\ref{fig:Op_Neel_HE_N=20}, we conclude that the discrepancy comes from the quantum device errors and noises rather than numerical errors of the first-order Trotterization. 
This also explains the disagreement between the values of quantum devices and the MPS-TDVP method, presented in Fig.~\ref{fig:Op_Neel_HE_large} for $N=96$ and $N=100$ with $H_{\mathrm{Dimer}}$.
Besides, this conclusion is also supported by the comparison of the results between $\texttt{ibm\_torino}$ and $\texttt{ibm\_sherbrooke}$ in Fig. \ref{fig:Op_Neel_HE_N=20}. The results of $\texttt{ibm\_torino}$ show certainly better accuracy than the results from $\texttt{ibm\_brisbane}$ and $\texttt{ibm\_sherbrooke}$ because $\texttt{ibm\_torino}$, $\texttt{ibm\_sherbrooke}$, and $\texttt{ibm\_sherbrooke}$ have $0.8 \%$, $1.7 \%$, and $1.9 \%$ EPLG (Error Per Layered Gate), respectively, in a chain of 100 qubits \cite{ibmq-system}. The EPLG measures the average gate process error in a layered chain of $100$ qubits. It is derived from a similar quantity known as layer fidelity (LF), and the LF is the process fidelity of the layered chain of $100$ qubits \cite{mckay2023benchmarking}.
Since the quantum circuit implementation for the $H_{\mathrm{Dimer}}$ consists of the two parts, $J_1$ terms and $J_2$ terms, we conjecture that the main reason for the discrepancy of the $H_{\mathrm{Dimer}}$ than $H_{\mathrm{iso}}$ is a longer circuit depth by the $J_2$ terms and, in particular, \texttt{CX} gates implementing the \texttt{SWAP} gates of the $J_2$ terms.

In our future work, we will study how to fine-tune the parameters of the error mitigation methods discussed in Sec.~\ref{sec:error_mitigations}. In particular, we assume that there will be a better extrapolation fitting function for the $H_{\mathrm{Dimer}}$ cases.
Additionally, we will explore other quantum error mitigation methods and find ways to combine them more efficiently. Furthermore, we plan to extend our study to include valid state preparation before the time evolution, allowing us to tackle more realistic quantum simulation problems.

\section{Conclusion and outlook}\label{sec:conclusion}

In conclusion, our study represents a significant step forward in the realm of quantum simulation before the fault tolerance quantum era, as we successfully implemented the quantum simulation of a frustrated quantum spin-$\frac{1}{2}$ antiferromagnetic Heisenberg spin chain on IBM's superconducting quantum computer. The incorporation of both nearest-neighbor $(J_1)$ and next-nearest-neighbor $(J_2)$ exchange interactions, particularly utilizing first-order Trotterization for the latter, demonstrates the versatility and capability of quantum computing technologies. Notably, our application of second-order Trotterization for the isotropic Heisenberg spin chain, coupled with precise measurements of staggered magnetization expectation values across a substantial range of qubits (up to 100), establishes the potential of these quantum devices for investigating properties of large-scale quantum systems.

The constant circuit depth achieved in each Trotter step, independent of the initial qubit number, adds a practical dimension to our findings, addressing a critical aspect of quantum simulation scalability. Moreover, our ability to accurately measure expectation values for such a large-scale quantum system using superconducting quantum computers underscores their utility in probing the intricacies of many-body quantum systems. 

In this study, we broaden the applicability of noisy quantum computers to encompass more intricate scenarios involving the time dynamics of Hamiltonians on a larger scale, specifically within the context of noisy superconducting quantum computers. 
In the future, our efforts pave the way for forthcoming quantum computing calculations, showcasing the quantum advantage over classical methods in simulating intricate quantum systems more prominently.
As we continue to push the boundaries of quantum computing capabilities, our findings contribute to the growing body of evidence supporting the transformative potential of quantum computers in advancing our understanding of quantum phenomena.

\section*{Acknowledgments}
We thank Taku Izubuchi for useful discussions and communications. T.A.C would also like to thank Kyoungchul Kong and John Ralston for the interesting discussions.
This research used quantum computing resources of the Oak Ridge Leadership Computing Facility, which is a DOE Office of Science User Facility supported under Contract DE-AC05-00OR22725. 
This research used resources of the National Energy Research Scientific Computing Center, a DOE Office of Science User Facility supported by the Office of Science of the U.S. Department of Energy under Contract No. DE-AC02-05CH11231 using NERSC award NERSC DDR-ERCAP0024165. R.S.S. is supported by the Special Postdoctoral Researchers Program of RIKEN and RIKEN-BNL Research Center.

\appendix
\section{Averaged staggered magnetization for $N=20$}\label{app:staggered}
In this section, we present the averaged value of the staggered magnetization with respect to the N\'eel state for simulation time in the case of isotropic Heisenberg Hamiltonian $H_{\mathrm{iso}}$ and the Dimer Hamiltonian $H_{\mathrm{Dimer}}$ presented in Fig. \ref{fig:Op_Neel_HB_N=20} and \ref{fig:Op_Neel_HE_N=20}. 
For the case of $H_{\mathrm{iso}}$ with $N=20$ qubits, we perform 5 repeated experiments of the staggered magnetization on \texttt{ibm$\_$sherbrooke} for both OBC and PBC, and obtain the corresponding averaged value in each Trotter time steps up to $t=4$.
The following term of $\pm$ represents the standard deviation of the data.

\begin{table}[h] 
\centering 
\renewcommand{\arraystretch}{1.3} 
\begin{tabular}{|>{\centering\arraybackslash}p{3cm}|>{\centering\arraybackslash}p{4.5cm}|>{\centering\arraybackslash}p{4.5cm}|} 
 \hline
 \multirow{2}{*}{\textbf{Time}, $t$} & \multicolumn{2}{c|}{\textbf{Staggered Magnetization}} \\ \cline{2-3}
  & \textbf{OBC} & \textbf{PBC} \\
  \hline
  0.5 & $ -0.3752 \pm 0.0011 $ &  $-0.3680 \pm 0.0012$  \\
 \hline
 1.0 & $ -0.1555 \pm 0.0004 $ & $-0.1448 \pm 0.0007$  \\
 \hline
 1.5 & $ 0.03741 \pm 0.00032 $ &  $0.591 \pm 0.0007$  \\
 \hline
 2.0 & $ 0.0811 \pm 0.0017 $  & $0.0836 \pm 0.0009 $  \\
 \hline
 2.5 & $ 0.0417 \pm 0.0008 $ &  $0.0413 \pm 0.0008 $  \\
 \hline
 3.0 & $ -0.0099 \pm 0.0005 $ &  $-0.0156 \pm 0.0010$  \\
 \hline
 3.5 & $ -0.0333 \pm 0.0006 $ & $-0.0303 \pm 0.0007$  \\
 \hline
 4.0 & $-0.0201 \pm 0.0008 $  & $-0.0137 \pm 0.0010$  \\
 \hline
\end{tabular}
\caption{Averaged staggered magnetization with time for $H_{\mathrm{iso}}$ and $N=20$ from 5 experiments on \texttt{ibm$\_$sherbrooke}. The $\pm$ terms represent the standard deviation of the data.}
\label{tab:HBN20}
\end{table}

In addition, for $H_{\mathrm{Dimer}}$ with $N=20$ qubits, we use \texttt{ibm$\_$torino} to measure the staggered magnetization repeatedly 5 times at each Trotter time step up to $t=1$, and determine the average value. 
The standard deviations in Table \ref{tab:HBN20} and \ref{tab:HEN20} show that each experimental data is distributed close to the average value. 

\begin{table}[h] 
\centering 
\renewcommand{\arraystretch}{1.3} 
\begin{tabular}{|>{\centering\arraybackslash}p{3cm}|>{\centering\arraybackslash}p{4.5cm}|>{\centering\arraybackslash}p{4.5cm}|} 
 \hline
 \multirow{2}{*}{\textbf{Time}, $t$} & \multicolumn{2}{c|}{\textbf{Staggered Magnetization}} \\ \cline{2-3}
  & \textbf{OBC} & \textbf{PBC} \\
  \hline
 0.2 & $-0.4665 \pm 0.0005$ & $-0.470 \pm 0.006$  \\
 \hline
 0.4 & $-0.4383 \pm 0.0014$  &  $-0.427 \pm 0.005$   \\
 \hline
 0.6 & $-0.3264 \pm 0.0012$  & $-0.320 \pm 0.004$  \\
 \hline
 0.8 & $-0.2252 \pm 0.0011 $  & $-0.2089 \pm 0.0030 $  \\
 \hline
 1.0 & $-0.1328 \pm 0.0005 $  & $-0.1245 \pm 0.0014 $   \\
 \hline
\end{tabular}
\caption{Averaged staggered magnetization with time for $H_{\mathrm{Dimer}}$ and $N=20$ from 5 experiments on \texttt{ibm$\_$torino}. The $\pm$ terms represent the standard deviation of the data.}
\label{tab:HEN20}
\end{table}

\section{Circuit depth and \texttt{CX} gate counts}\label{app:circuitdepth}
We present the circuit depth and \texttt{CX} gate counts for Trotter steps associated with the Trotterization circuits for the Hamiltonians $H_{\mathrm{iso}}$ and $H_{\mathrm{Dimer}}$, respectively. The circuit depth is an important measure of how many operations one can implement before the coherence breaks down in a quantum computer. Therefore, the circuit depth associated with the Trotterization circuits essentially captures how reliable the time evolutions of spin-chain under the isotropic Heisenberg and the Dimer Hamiltonians are, and how the system size scales with the initial number of qubits.
In addition, \texttt{CX} is the noisiest gate in the base gate set. Hence, measuring the number of \texttt{CX} gates in a circuit can be used to estimate the noise of the circuit.
In the following tables, the circuit depths and the number of \texttt{CX} gates are measured transpiling with the optimization level 3 in Qiskit.

\begin{table}[h!]
    \centering
\begin{tabular}{|>{\centering\arraybackslash}p{3cm}|>{\centering\arraybackslash}p{2.5cm}|>{\centering\arraybackslash}p{2.5cm}|>{\centering\arraybackslash}p{2.5cm}|>{\centering\arraybackslash}p{2.5cm}|}
    \hline
  \multirow{3}{*}{\textbf{Trotter Step}} & \multicolumn{4}{|c|}{\textbf{Circuit depth for $H_{\mathrm{iso}}$}}  \\ \cline{2-5}
        & \multicolumn{2}{|c|}{\textbf{OBC}} & \multicolumn{2}{|c|}{\textbf{PBC}} \\ \cline{2-5}
        & \multicolumn{1}{|c|}{$\textbf{N=20}$} & \multicolumn{1}{|c|}{$\textbf{N=100}$} & \multicolumn{1}{|c|}{$\textbf{N=20}$} & \multicolumn{1}{|c|}{$\textbf{N=96}$}  \\ \hline
        \textbf{1} & 41 & 41 & 40 & 47 \\
        \hline
        \textbf{2} & 67 & 67 & 66 & 77 \\        
        \hline
        \textbf{3} & 93 & 93 & 92 & 106 \\ 
        \hline
        \textbf{4} & 119 & 119 & 118 & 135 \\ 
        \hline
        \textbf{5} & 145 & 145 & 144 & 164 \\ 
        \hline
        \textbf{6} & 171 & 171 & 170 & 193 \\ 
        \hline
        \textbf{7} & 197 & 197 & 196 & 222 \\ 
        \hline
        \textbf{8} & 223 & 223 & 222 & 251 \\ \hline
\end{tabular}
\caption{Circuit depth with respect to the Trotter steps for $H_{\mathrm{iso}}$ with $N=20, 96, 100$ qubits after transpiling with the optimization level 3.}
    \label{tab:circuit-depth-HB}
\end{table}

\begin{table}[h!]
    \centering
\begin{tabular}{|>{\centering\arraybackslash}p{3cm}|>{\centering\arraybackslash}p{2.5cm}|>{\centering\arraybackslash}p{2.5cm}|>{\centering\arraybackslash}p{2.5cm}|>{\centering\arraybackslash}p{2.5cm}|}
    \hline
  \multirow{3}{*}{\textbf{Trotter Step}} & \multicolumn{4}{|c|}{\textbf{No. of \texttt{CX} gates for $H_{\mathrm{iso}}$}}  \\ \cline{2-5}
        & \multicolumn{2}{|c|}{\textbf{OBC}} & \multicolumn{2}{|c|}{\textbf{PBC}} \\ \cline{2-5}
        & \multicolumn{1}{|c|}{$\textbf{N=20}$} & \multicolumn{1}{|c|}{$\textbf{N=100}$} & \multicolumn{1}{|c|}{$\textbf{N=20}$} & \multicolumn{1}{|c|}{$\textbf{N=96}$}  \\ \hline
        \textbf{1} & 87 & 447 & 90 & 432 \\
        \hline
        \textbf{2} & 144 & 744 & 150 & 720 \\        
        \hline
        \textbf{3} & 201 & 1041 & 210 & 1008 \\ 
        \hline
        \textbf{4} & 258 & 1338 & 270 & 1296 \\ 
        \hline
        \textbf{5} & 315 & 1635 & 330 & 1584 \\ 
        \hline
        \textbf{6} & 372 & 1932 & 390 & 1872 \\ 
        \hline
        \textbf{7} & 429 & 2229 & 450 & 2160 \\ 
        \hline
        \textbf{8} & 486 & 2526 & 510 & 2448 \\ \hline
\end{tabular}
\caption{Number of \texttt{CX} gates for $H_{\mathrm{iso}}$ after transpiling with the optimization level 3.}
    \label{tab:cxgates-HB}
\end{table}

\begin{table}[h!]
    \centering
    \renewcommand{\arraystretch}{1.5} 
    \begin{tabular}{|c|*{8}{>{\centering\arraybackslash}p{1.5cm}|}} 
        \hline
        \multirow{4}{*}{\shortstack{\textbf{Trotter} \\ \textbf{step}}} & \multicolumn{8}{c|}{\textbf{Circuit depth for $H_{\mathrm{Dimer}}$}} \\ \cline{2-9}
        & \multicolumn{4}{c|}{\textbf{OBC}} & \multicolumn{4}{c|}{\textbf{PBC}} \\  \cline{2-9}
        & \multicolumn{2}{c|}{$\mathbf{N = 20}$} & \multicolumn{2}{c|}{$\textbf{N = 100}$} & \multicolumn{2}{c|}{$\textbf{N = 20}$} & \multicolumn{2}{c|}{$\textbf{N = 96}$} \\ \cline{2-9}
        & \multicolumn{1}{c|}{\shortstack{$\texttt{ibm} \_ $ \\ \texttt{sherbrooke}}} & \multicolumn{1}{c|}{\shortstack{$\texttt{ibm} \_ $ \\ \texttt{torino}}} & \multicolumn{1}{c|}{\shortstack{$\texttt{ibm} \_ $ \\ \texttt{brisbane}}} & \multicolumn{1}{c|}{\shortstack{$\texttt{ibm} \_ $ \\ \texttt{torino}}} & \multicolumn{1}{c|}{\shortstack{$\texttt{ibm} \_ $ \\ \texttt{sherbrooke}}} & \multicolumn{1}{c|}{\shortstack{$\texttt{ibm} \_ $ \\ \texttt{torino}}} & \multicolumn{1}{c|}{\shortstack{$\texttt{ibm} \_ $ \\ \texttt{sherbrooke}}} & \multicolumn{1}{c|}{\shortstack{$\texttt{ibm} \_ $ \\ \texttt{torino}}}  \\ \cline{2-9} 
        \hline
        \textbf{1} & 66  & 64  & 71  & 70  & 68  & 64 & 80  & 69  \\ \cline{2-9}
        \hline
        \textbf{2} & 141  & 132  & 155  & 144  & 144  & 132  & 163  & 146  \\ \cline{2-9}
        \hline
        \textbf{3} & 215  & 200  & 239  & 218  & 218  & 200  & 246  & 223  \\ \cline{2-9}
        \hline
        \textbf{4} & 298  & 268  & 293  & 292  & 292  & 268  & 329  & 300  \\ \cline{2-9}
        \hline
        \textbf{5} & 363  & 336  & 407  & 366  & 366  & 336  & 412  & 377  \\ \cline{2-9}
        \hline
    \end{tabular}
     \caption{Circuit depth with respect to the Trotter steps for $H_{\mathrm{Dimer}}$ with $N=20, 96, 100$ qubits after transpiling with the optimization level 3.}
    \label{tab:circuit-depth-HE}
\end{table}

\begin{table}[h!]
    \centering
    \renewcommand{\arraystretch}{1.5} 
    \begin{tabular}{|c|*{8}{>{\centering\arraybackslash}p{1.5cm}|}} 
        \hline
        \multirow{4}{*}{\shortstack{\textbf{Trotter} \\ \textbf{step}}} & \multicolumn{8}{c|}{\textbf{No. of \texttt{CX} gates for $H_{\mathrm{Dimer}}$}} \\ \cline{2-9}
        & \multicolumn{4}{c|}{\textbf{OBC}} & \multicolumn{4}{c|}{\textbf{PBC}} \\  \cline{2-9}
        & \multicolumn{2}{c|}{$\mathbf{N = 20}$} & \multicolumn{2}{c|}{$\textbf{N = 100}$} & \multicolumn{2}{c|}{$\textbf{N = 20}$} & \multicolumn{2}{c|}{$\textbf{N = 96}$} \\ \cline{2-9}
        & \multicolumn{1}{c|}{\shortstack{$\texttt{ibm} \_ $ \\ \texttt{sherbrooke}}} & \multicolumn{1}{c|}{\shortstack{$\texttt{ibm} \_ $ \\ \texttt{torino}}} & \multicolumn{1}{c|}{\shortstack{$\texttt{ibm} \_ $ \\ \texttt{brisbane}}} & \multicolumn{1}{c|}{\shortstack{$\texttt{ibm} \_ $ \\ \texttt{torino}}} & \multicolumn{1}{c|}{\shortstack{$\texttt{ibm} \_ $ \\ \texttt{sherbrooke}}} & \multicolumn{1}{c|}{\shortstack{$\texttt{ibm} \_ $ \\ \texttt{torino}}} & \multicolumn{1}{c|}{\shortstack{$\texttt{ibm} \_ $ \\ \texttt{sherbrooke}}} & \multicolumn{1}{c|}{\shortstack{$\texttt{ibm} \_ $ \\ \texttt{torino}}}  \\ \cline{2-9} 
        \hline
        \textbf{1} & 138  & 138  & 738  & 738  & 150 & 150  & 720 & 720  \\ \cline{2-9}
        \hline
        \textbf{2} & 288   & 288  & 1548  & 1548  & 315 & 315  & 1512 & 1512  \\ \cline{2-9}
        \hline
        \textbf{3} & 438 & 438  & 2358 & 2358  & 480 & 480  & 2304 & 2304  \\ \cline{2-9}
        \hline
        \textbf{4} & 588 & 588  & 3168 & 3168  & 645 & 645  & 3096 & 3096  \\ \cline{2-9}
        \hline
        \textbf{5} & 738 & 738  & 3978 & 3978  & 810 & 810  & 3888 & 3888  \\ \cline{2-9}
        \hline
    \end{tabular}
    \caption{Number of \texttt{CX} gates for $H_{\mathrm{Dimer}}$ after transpiling with the optimization level 3.}
    \label{tab:cxgates-HE}
\end{table}

\section{Circuit qubit mapping layout}\label{app:qubit_mapping}
\begin{figure}[h]
\centering
  \subfloat[$\texttt{ibm\_sherbrooke}$, 20 qubits, PBC]{%
    \includegraphics[width=0.48\textwidth]{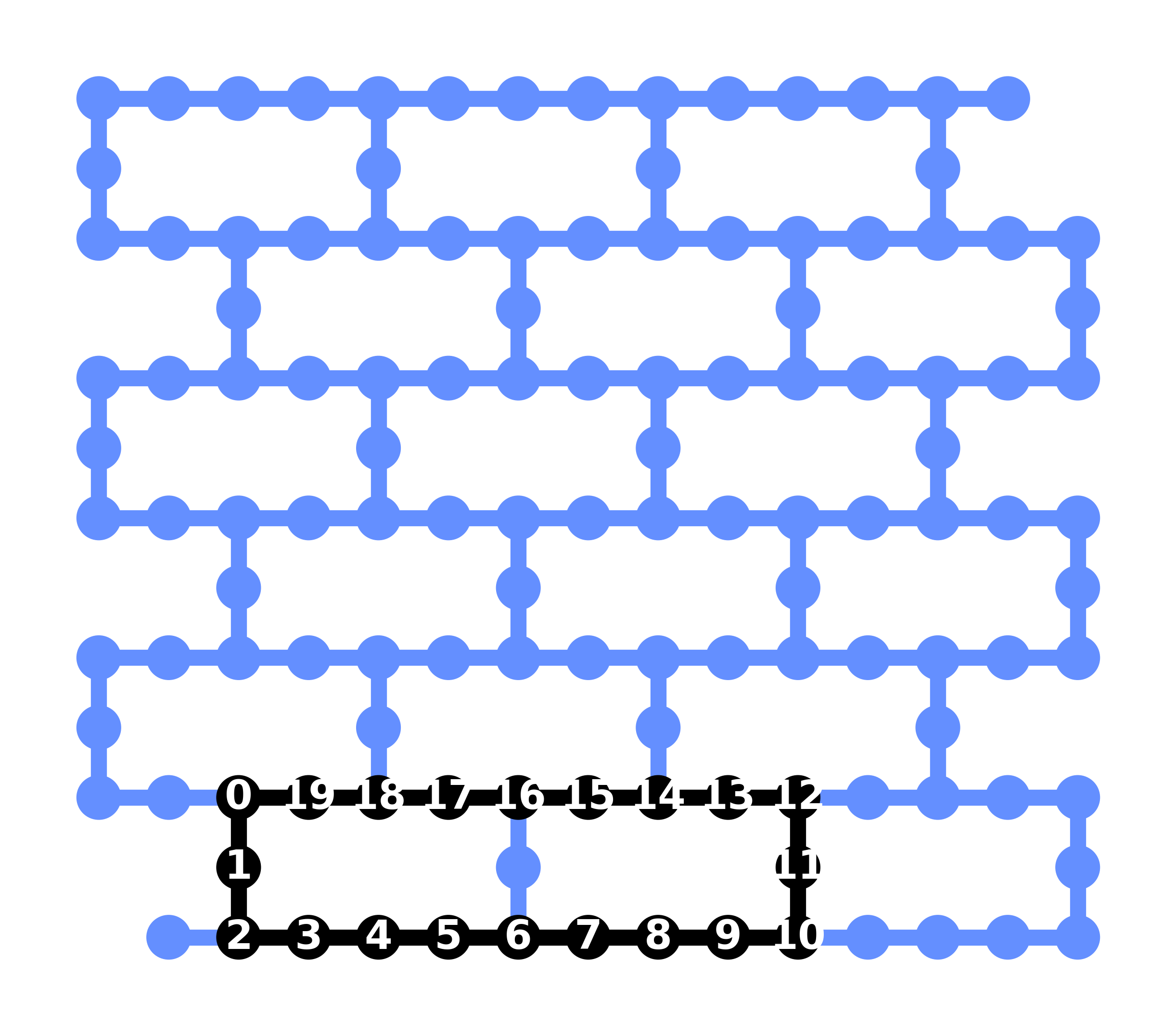}}
  ~
  \subfloat[$\texttt{ibm\_sherbrooke}$, 20 qubits, OBC]{%
    \includegraphics[width=0.48\textwidth]{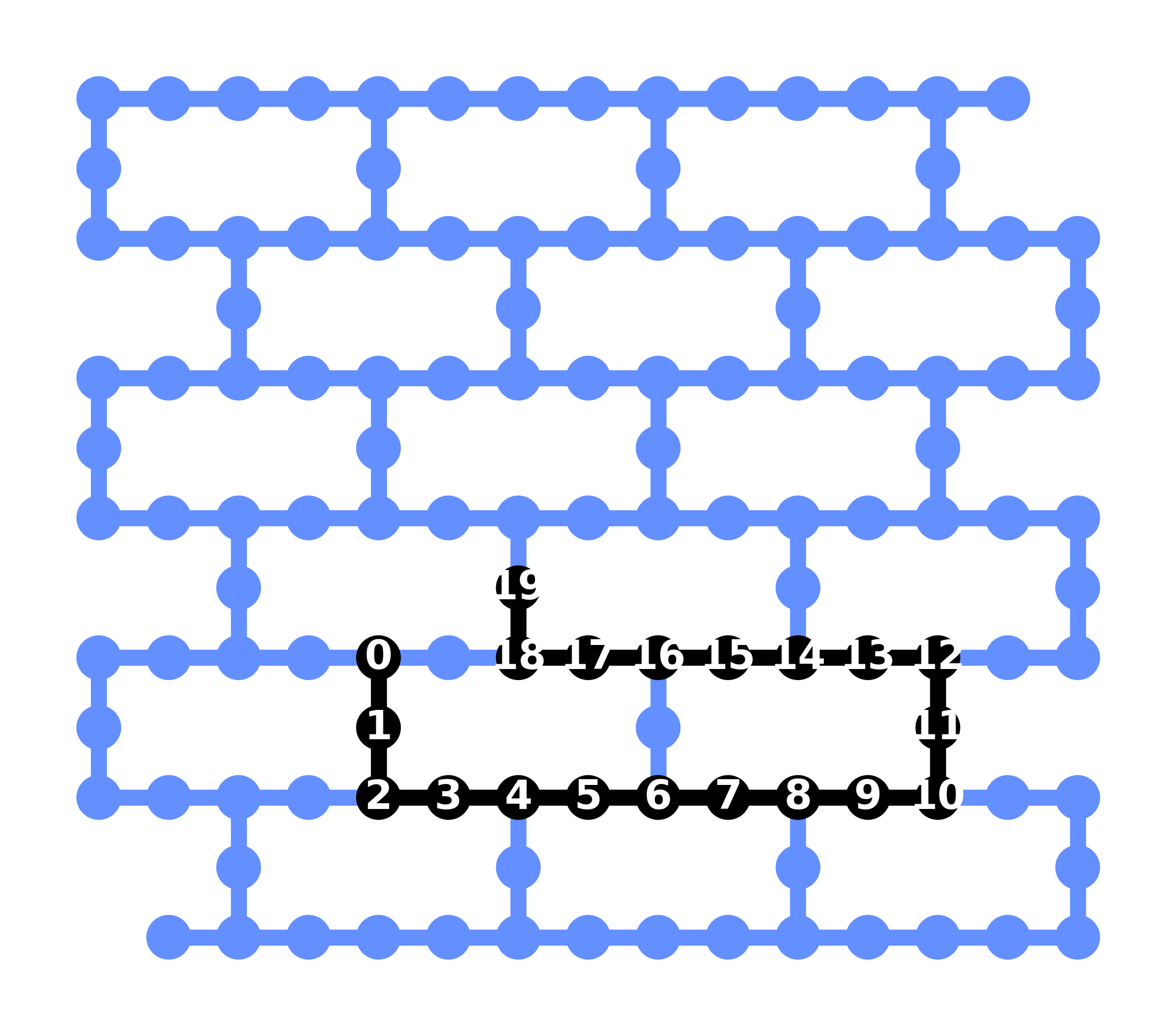}}\\
  \subfloat[$\texttt{ibm\_sherbrooke}$, 96 qubits, PBC]{%
    \includegraphics[width=0.48\textwidth]{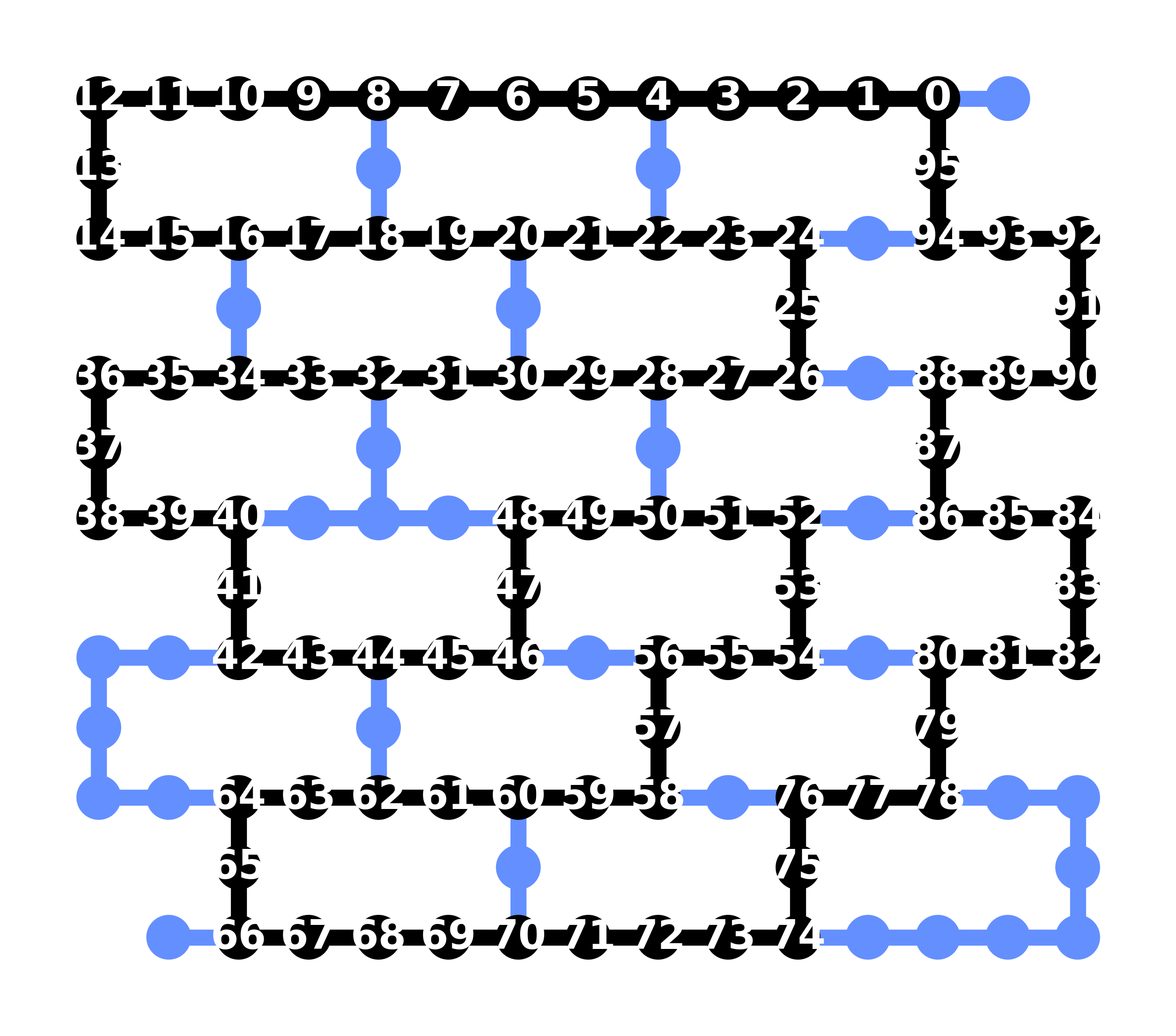}}
  ~
  \subfloat[$\texttt{ibm\_brisbane}$, 100 qubits, OBC]{%
    \includegraphics[width=0.48\textwidth]{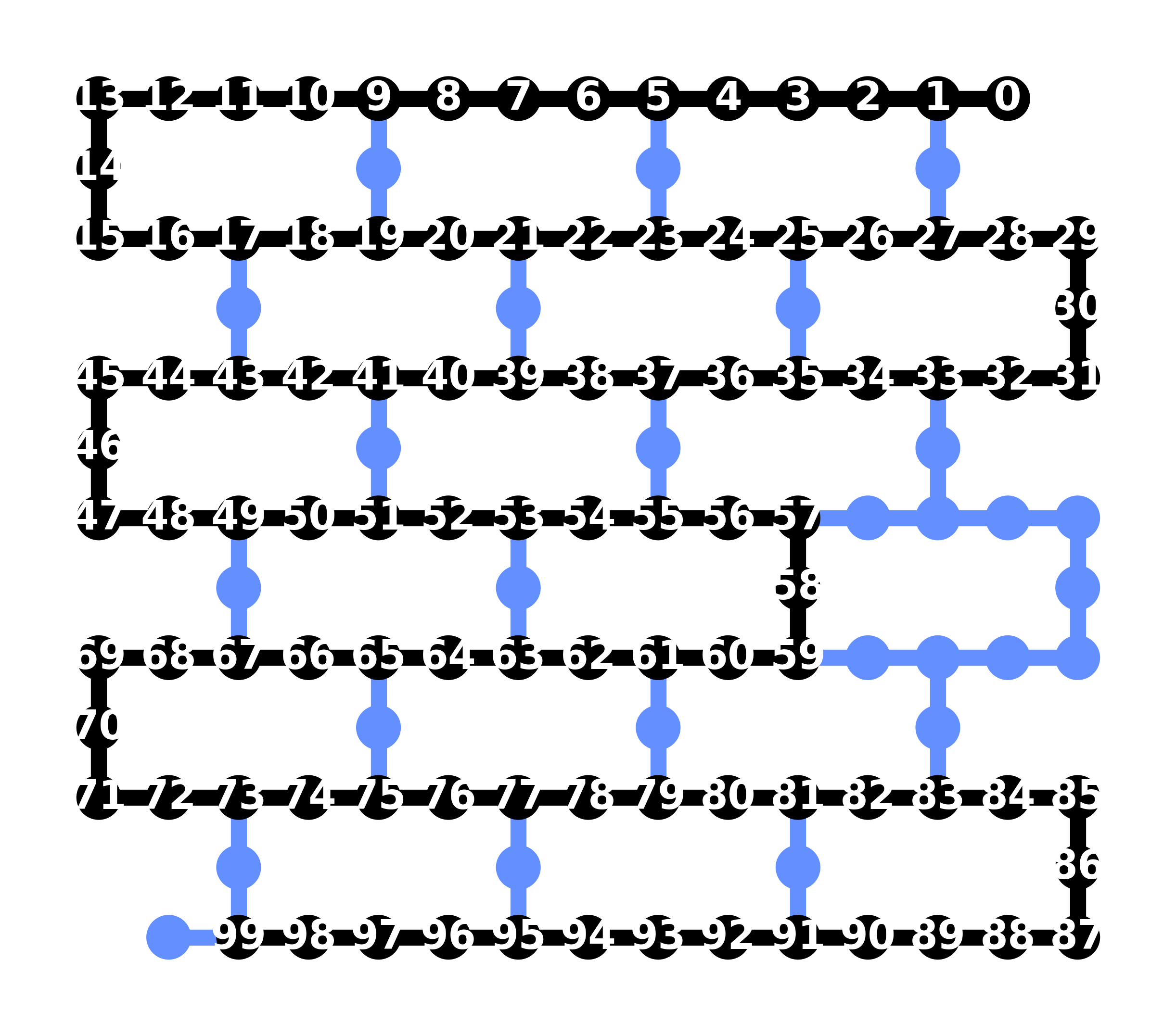}}
    
  \vspace{0.5cm}
  \caption{Circuit qubit mapping layout.}
  \label{fig:QMapping}
\end{figure}

Here, we present the qubit mappings of \texttt{ibm$\_$sherbrooke} and \texttt{ibm$\_$brisbane} for $N=20, 96$ and 100 qubits in Fig.~\ref{fig:QMapping} which are used in our experiments.

\pagebreak
\newpage

\bibliography{references}

\end{document}